\documentclass[aps,prl,superscriptaddress,longbibliography,twocolumn]{revtex4-2}
\usepackage[english]{babel}
\usepackage{amsmath,amssymb,bbm,graphicx,color,comment}
\usepackage[bookmarks=true,colorlinks,citecolor=blue,urlcolor=blue]{hyperref}

\usepackage{dsfont}
\usepackage{soul}
\usepackage{subfigure}
\usepackage{extarrows}
\usepackage{float}
\usepackage{txfonts}

\usepackage{mathtools}
\usepackage[usenames,dvipsnames]{xcolor} 
\usepackage[export]{adjustbox}
\usepackage{adjustbox}

\usepackage{dcolumn}   
\usepackage{bm}        
\usepackage{amsfonts}  
\usepackage{filecontents}
\usepackage{lineno}

\usepackage{pifont}
\newcommand{\cmark}{\ding{51}}%
\newcommand{\xmark}{\ding{55}}%

\renewcommand{\vec}[1]{\bm{#1}}

\usepackage{times}

\begin{document}

\title{Spontaneous Crystal Thermal Hall Effect in Insulating Altermagnets}%

\author{Rhea Hoyer}
\affiliation{Department of Physics, Johannes Gutenberg University Mainz, 55128 Mainz, Germany}

\author{Rodrigo Jaeschke-Ubiergo}
\affiliation{Department of Physics, Johannes Gutenberg University Mainz, 55128 Mainz, Germany}

\author{Kyo-Hoon Ahn}
\affiliation{Institute of Physics, Czech Academy of Sciences, Cukrovarnická 10, 162 00 Praha 6, Czech Republic}

\author{Libor Šmejkal}
\affiliation{Department of Physics, Johannes Gutenberg University Mainz, 55128 Mainz, Germany}
\affiliation{Institute of Physics, Czech Academy of Sciences, Cukrovarnická 10, 162 00 Praha 6, Czech Republic}

\author{Alexander Mook}
\affiliation{Department of Physics, Johannes Gutenberg University Mainz, 55128 Mainz, Germany}

\begin{abstract}
    We show that magnetic insulators with a collinear and compensated order can exhibit a thermal Hall effect even at zero magnetic field if they have altermagnetic symmetries.
    We predict a finite thermal Hall conductivity vector $\bm{\kappa}_\text{H}$ for a rutile-inspired effective spin model with Dzyaloshinskii-Moriya interaction. Within the linear spin-wave theory, we identify two magnon branches that carry identical Berry curvature and give rise to a finite $\bm{\kappa}_\text{H}$, which can be controlled by the Néel vector orientation and by strain. The thermal Hall response is further complemented with a spin Nernst response to contrast spin and heat transport in altermagnetic insulators with those in ferromagnets and antiferromagnets. Our results establish the crystal thermal Hall effect of magnons and we discuss material candidates for experimental realization, such as MnF$_2$, CoF$_2$, and NiF$_2$. 
\end{abstract}

\date{\today}%
\maketitle
Antiferromagnets---or, more broadly, magnets with \emph{compensated} magnetic order---keep receiving attention in quantum materials science \cite{Shen2008, Plakida2010, Yuan2022quantummagnonics} and spintronics \cite{Jungwirth2016, Baltz2018, bonbien2021topological, Xiong2022, Han2023}. They are an abundant low-temperature phase in, e.g., high-$T_\text{c}$ superconductor parent compounds \cite{Shen2008, Plakida2010, Keimer2015}, proximate quantum spin-liquids in Kitaev magnets \cite{Savary2016, Trebst2022}, topological spintronics states \cite{Smejkal2018}, and van der Waals magnets \cite{Wang2022}, and have promising properties for technological applications \cite{Jungwirth2016}. 
Alas, as many of them are insulators, conventional charge probes yield null information. Instead, the thermal Hall effect (THE) provides a window into charge-neutral emergent excitations above the many-body ground state \cite{Li2020review, Guo2022, Zhang2023reviewTHE}. The THE describes the observation that a temperature gradient $\vec{\nabla} T$ causes a heat current density $\vec{q} = \vec{\kappa}_\text{H} \times \vec{\nabla} T$ in transverse direction, where $\vec{\kappa}_\text{H} = (\kappa_{yz}, \kappa_{zx}, \kappa_{xy})^\text{T}$ is the thermal Hall conductivity vector. 
Since charge-neutral excitations such as phonons and magnetic excitations carry energy, they can contribute to $\vec{\kappa}_\text{H}$ \cite{Strohm2005, Kitaev2006, Sheng2006, Kagan2008,Katsura2010, Onose2010, Matsumoto2011PRL, Matsumoto2011, Ideue2012, Mori2014, Hirschberger2015, Romhanyi2015, Savary2016, watanabe2016, Ideue2017, McClarty2017, Doki2018, Kasahara2018, Kasahara2018PRL, Grissonnanche2019, Kawano2019, Zhang2020SU3, Li2020, Chen2020Ferroelectrics, Akazawa2020, Zhang2021, Chen2022THE, Neumann2022, Guo2022, Zhang2023reviewTHE, Czajka2022, Mangeolle2022, Li2023PRB, Mangeolle2022}. Thus, understanding the symmetry constraints of the THE is a key prerequisite for interpreting transport data and drawing conclusions on the ground state \cite{Zhang2023reviewTHE}.
A central question is whether the compensated magnetic order gives rise to a spontaneous \emph{anomalous} THE at zero magnetic field. While the necessary time-reversal symmetry breaking is obviously the case for uncompensated magnets, it also occurs in certain \emph{noncollinear} compensated magnets \cite{Mook2019Coplanar}. For \emph{collinear} compensated magnets, however, prior studies have relied on magnetic fields to explicitly break time-reversal symmetry and to generate $\vec{\kappa}_\text{H}$ \cite{ Hirschberger2015, Kawano2019, Zhang2020SU3, Chen2022THE, Neumann2022, Mangeolle2022, Li2023PRB}. 

Here, we ask the question: \emph{Can insulating collinear magnets with a compensated order exhibit a THE at zero magnetic field?}
Valuable intuition can be gained from the theory of altermagnetism \cite{Smejkal2022altermagnets, Smejkal2022emergent}. Altermagnetic order is compensated and collinear, and characterized by an unconventional $d$-, $g$- or $i$-wave spin polarized order \cite{Smejkal2022altermagnets, Smejkal2022emergent}, and corresponding spin splitting which was recently experimentally confirmed \cite{LeeSmejkal2024, KrempaskySmejkal2024, Reimers2024}. Altermagnets thus lack genuine antiferromagnetic symmetries that map the sublattices onto each other by inversion or translation and lead to a Kramers band degeneracy. Since these symmetries are what rule out time-reversal odd phenomena in antiferromagnets, altermagnets---whose sublattices are instead related by rotation---can exhibit Hall-type transport \cite{Smejkal2022altermagnets, Smejkal2022emergent, Smejkal2022AHEANTI}. A finite anomalous Hall effect \cite{Smejkal2020CrystalHall, Samanta2020, Mazin2021, Feng2022, Reichlova2021, Tschirner2023} and its thermal counterparts \cite{ZhouPRL2024} have been identified in \emph{metallic} altermagnets. These effects are also referred to as the \textit{crystal} Hall effects to emphasize \textit{firstly}, that they arise because the crystalline structure breaks the antiferromagnetic sublattice mapping, and \textit{secondly}, their strong angular dependence on the Néel vector orientation due to the extra order parameters that originate from the interplay between crystallographic and magnetic structure \cite{Smejkal2020CrystalHall}.

Given the above context, we rephrase our question: \emph{Is there a crystal THE in insulating altermagnets?} We give an affirmative answer by providing a transparent toy model, considering magnons as microscopic heat carriers, since they are directly affected by the altermagnetic symmetry breaking \cite{SmejkalMagnons23, Gohlke2023PRL, CuiPRB2023}. 
We show that $\vec{\kappa}_\text{H}$ arises from the spin-orbit coupling in the form of Dzyaloshinskii-Moriya interaction (DMI) \cite{Dzyaloshinsky58, Moriya60} accounting for the local crystallographic environment of the magnetic ions. Our toy model supports two magnon bands with identical Berry curvature and, as a result, we find a finite $\vec{\kappa}_\text{H}$ that shows a strong dependence on the N\'{e}el vector orientation. The influence of symmetry is further emphasized by demonstrating that strain can break symmetries and turn $\vec{\kappa}_\text{H}$ on and off. 
We also study the relativistic spin Nernst effect (SNE) and show that it relies on the counterplay between the spin-orbit coupling and the altermagnetic band splitting. 
Our results suggest that the THE can be a valuable probe of altermagnetism in insulating magnets. 

\begin{figure}[]
       \includegraphics[width=\columnwidth]{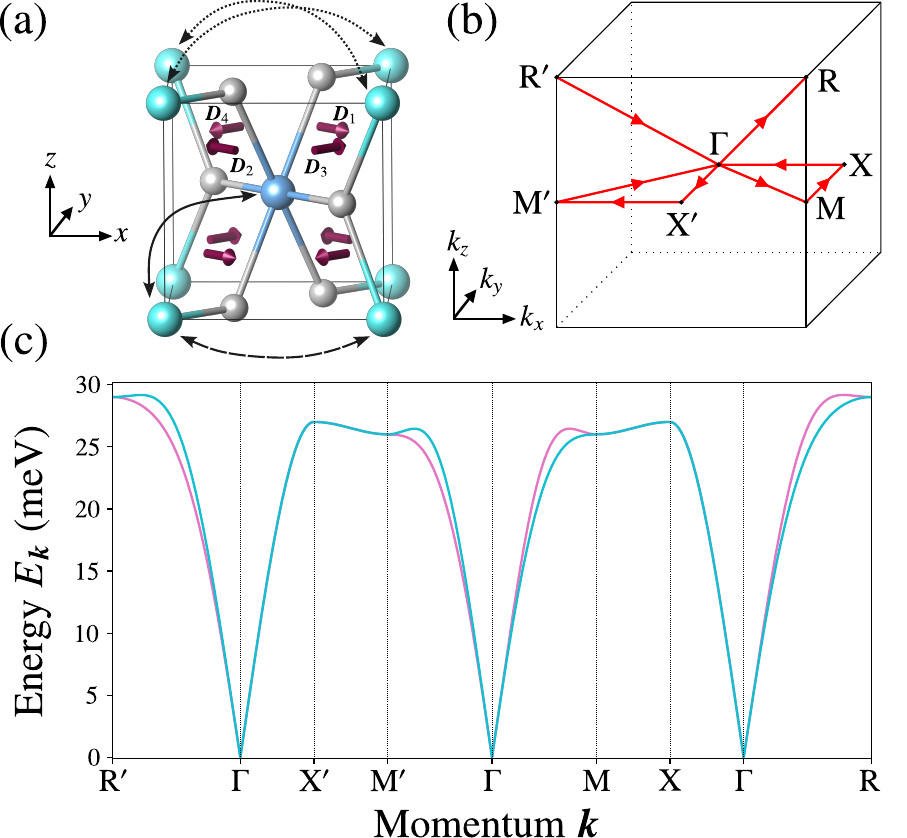}
       \caption{Altermagnetic spin model. (a) Unit cell with magnetic/nonmagnetic atoms (blue/white) and DMI 
       vectors $\vec{D}_i$ (red). The solid/dashed/dotted arrows indicate the exchange interactions $J_1$/$J_2$/$J_3 \pm \Delta$. (b) High-symmetry path in the Brillouin zone. (c) Magnon dispersion with $S = 5/2$, $J_1 = 1.0$ meV, $J_2 = -0.3$ meV, $J_3 = -0.2$ meV, $\Delta = D = 0.1$ meV, and the Néel vector $\bm{N} \parallel \hat{\vec{x}}$. Cyan (magenta) corresponds to a magnonic spin moment expectation value $\langle \vec{s} \rangle = + \hbar \hat{\vec{N}}$ $(-\hbar \hat{\vec{N}})$.}
       \label{fig:1}
\end{figure}

\paragraph{Model.}
We build a minimal model of a two-sublattice altermagnet by analogy with the rutile crystal structure, as shown in Fig.~\ref{fig:1}(a), but, without loss of generality, we consider a cubic unit cell with a lattice constant of $a = 0.5\,$nm. The spin Hamiltonian is given by
\begin{align}
\label{eq:spin-Hamiltonian}
    H_{\text{s}} =& \sum_{r=1}^3 \sum_{\langle i, j \rangle_r} J_{ij} \bm{S}_{i} \cdot \bm{S}_{j} 
    + \sum_{\langle i, j \rangle } \bm{D}_{ij} \cdot  \left( \bm{S}_{i} \times \bm{S}_{j} \right).
\end{align}
Here, $J_{ij} = J_r$ denotes Heisenberg exchange that we include up to the third-nearest neighbor shell; shells are labeled by $r$. The different exchange paths are indicated by arrows in Fig.~\ref{fig:1}(a). For nearest and second neighbors (solid and dashed arrow), we set $J_1 > 0$ and $J_2<0$, respectively. Due to the nonmagnetic ions, there are two crystallographically inequivalent paths for third neighbors (dotted arrows) \cite{SmejkalMagnons23, Gohlke2023PRL}, parametrized by $J_3 \pm \Delta$, with $J_3<0$. Relativistic effects are included by a nearest-neighbor DMI \cite{Dzyaloshinsky58, Moriya60}, which is allowed by the DMI rules. The DMI vectors $\vec{D}_{ij}$ are indicated in dark red in Fig.~\ref{fig:1}(a), following the convention that site $i$ is the central ion and site $j$ one of the corner ions; they read 
$
    \bm{D}_{1} = - \bm{D}_{2} = D(1,-1,0)^\text{T}/\sqrt{2}
$
and 
$
    \bm{D}_{3} = - \bm{D}_{4} = D (1,1,0)^\text{T} / \sqrt{2}
$.

The classical ground state of $H_{\text{s}}$ is the collinear N\'{e}el order, with the order parameter $\vec{N} = \vec{M}_\text{A} - \vec{M}_\text{B}$ being the difference of the sublattice magnetizations, $\vec{M}_\text{A}$ and $\vec{M}_\text{B}$, on the A and B sublattice. The signs of the $J$'s are chosen to avoid exchange frustration. The collinear order persists for $D \ne 0$ because $\sum_{i=1}^4 \bm{D}_i = 0$, and we find no classical weak ferromagnetic moment. We emphasize that the rutiles can be compatible with weak ferromagnetism \cite{dzialoshinskii1958magnetic}, but $H_{\text{s}}$ deliberately does not include all symmetry-allowed magnetic interactions---it lacks, in particular, local anisotropies \cite{Moriya1960NiF2}---to keep the discussion as simple as possible. 

To perform a spin-wave analysis by expanding in fluctuations around the collinear order we rewrite the spins as 
$
    \vec{S}_i^{\text{A}/\text{B}}
    =
    \widetilde{S}^{\text{A}/\text{B}, x}_i
    \hat{\vec{x}}
    +
    \widetilde{S}^{\text{A}/\text{B}, y}_i 
    \hat{\vec{y}}
    +
    \widetilde{S}^{\text{A}/\text{B}, z}_i 
    \hat{\vec{z}}
$,
where $\{ \hat{\vec{x}}, \hat{\vec{y}}, \hat{\vec{z}} \}$ form a local coordinate system, with $\hat{\vec{z}} = \hat{\vec{N}}= \vec{N} / |\vec{N}|$. After a Holstein-Primakoff transformation, the spins read 
$\widetilde{S}^{\text{A}, z}_i = S - a^\dagger_i a_i$, $\widetilde{S}^{\text{A}, x}_i - \mathrm{i} \widetilde{S}^{\text{A}, y}_i = a^\dagger_i \sqrt{2S - a^\dagger_i a_i}$, $\widetilde{S}^{\text{B}, z}_i = - S + b^\dagger_i b_i$ and $\widetilde{S}^{\text{B}, x}_i - \mathrm{i} \widetilde{S}^{\text{B}, y}_i = \sqrt{2S - b^\dagger_i b_i} \,b_i $, 
\cite{holsteinprimakoff1940}, where $S$ is the spin length and the $a^\dagger_i$/$b^\dagger_i$'s ($a_i$/$b_i$'s) are bosonic creation (annihilation) operators on the A/B sublattice. The spin Hamiltonian is expanded in the number of bosons (or $1/\sqrt{S})$, $H_\text{s} = H_0 + H_1 + H_2 + H_3 + H_4 + \ldots$, where $H_0$ is the ground state energy, $H_1 = 0$ because we expand around a classically stable order, $H_2$ is the bilinear part, and $H_3$ and beyond capture magnon-magnon interactions. The full details can be found in the Supplemental Material (SM) \cite{SM}. After a Fourier transformation of the bosonic operators, 
$
	a_{i}/b_{i} = \frac{1}{\sqrt{N}} \sum_{\bm{k}} \mathrm{e}^{\mathrm{i} \bm{k} \cdot \bm{r}_{i}} a_{\bm{k}}/b_{\bm{k}}
$,
a block-diagonal bilinear Hamiltonian is found,
$
    H_2 
        = 
        \frac{1}{2} \sum_{\vec{k}} \sum_{\sigma = \pm}
            \vec{\Psi}_{\vec{k},\sigma}^\dagger H_{\vec{k},\sigma} \vec{\Psi}_{\vec{k},\sigma}.
$
The two spinors are given by
$
    \vec{\Psi}_{\vec{k},+}^\dagger
    =
    ( a^\dagger_{\vec{k}}, b_{-\vec{k}} ) 
$
and
$
    \vec{\Psi}_{\vec{k},-}^\dagger
    =
    ( b^\dagger_{\vec{k}}, a_{-\vec{k}} )
$,
and the respective block kernels read
\begin{align}
    H_{\vec{k},\pm}
    &=
    S
    \begin{pmatrix}
        A_{\vec{k}} \pm \Delta_{\vec{k}} & B_{\vec{k}} - \mathrm{i} D_{\vec{k}} \\
        B_{\vec{k}} + \mathrm{i} D_{\vec{k}} & A_{\vec{k}} \mp \Delta_{\vec{k}} 
    \end{pmatrix}, 
\end{align}
where
\begin{subequations}
\begin{align}
    A_{\vec{k}} 
    &= 
    8J_1 - 6J_2 - 4J_3 + 2J_2\left( \cos k_x + \cos k_y + \cos k_z \right) \nonumber \\ 
    &\quad +2 J_3 \left[ \cos \left(k_x + k_y \right) + \cos \left(k_x - k_y \right) \right] ,
    \\
    \Delta_{\vec{k}} 
    &= 
    2\Delta \left[ \cos \left(k_x + k_y \right) - \cos \left(k_x - k_y \right) \right],
    \\
    B_{\vec{k}} 
    &= 
    2J_1 \left( \cos \frac{k_x+k_y+k_z}{2} + \cos \frac{k_x+k_y-k_z}{2} \right. \nonumber \\
    &\quad \left. + \cos \frac{k_x-k_y+k_z}{2} + \cos \frac{-k_x+k_y+k_z}{2} \right),
    \\
    D_{\vec{k}} 
    &= 
    2 D \hat{\vec{N}} \cdot \left( \hat{\vec{D}}_1 \cos \frac{k_x+k_y+k_z}{2} + \hat{\vec{D}}_2 \cos \frac{k_x+k_y-k_z}{2} \right. \nonumber \\
    &\quad  \left. + \hat{\vec{D}}_3 \cos \frac{k_x-k_y+k_z}{2} + \hat{\vec{D}}_4 \cos \frac{-k_x+k_y+k_z}{2} \right). \label{eq:DMIinH2}
\end{align}
\end{subequations}

We can perform the following calculations for each block separately. First, we Bogoliubov diagonalize the blocks, 
\begin{align}
    T^\dagger_{\vec{k}} H_{\vec{k},+} T_{\vec{k}} 
    &= 
    \text{diag}\left( E_{\vec{k},\alpha} , E_{-\vec{k},\beta} \right),
    \quad
    \begin{pmatrix}
        \alpha_{\vec{k}} \\ \beta^\dagger_{-\vec{k}}
    \end{pmatrix}
    =
    T_{\vec{k}}^{-1} \vec{\Psi}_{\vec{k},+},
    \nonumber \\
    T^\dagger_{\vec{k}} H_{\vec{k},-} T_{\vec{k}} 
    &= 
    \text{diag}\left( E_{\vec{k},\beta} , E_{-\vec{k},\alpha} \right), 
    \quad
    \begin{pmatrix}
        \beta_{\vec{k}} \\ \alpha^\dagger_{-\vec{k}}
    \end{pmatrix}
    =
    T_{\vec{k}}^{-1} \vec{\Psi}_{\vec{k},-}, 
\end{align}
so that
$
    H_2
        = 
        \sum_{\vec{k}} [
        E_{\vec{k},\alpha} ( \alpha^\dagger_{\vec{k}} \alpha_{\vec{k}} + \frac{1}{2} )
        +
        E_{\vec{k},\beta} ( \beta^\dagger_{\vec{k}} \beta_{\vec{k}} + \frac{1}{2} )
        ],
$
where $\alpha^\dagger_{\vec{k}}$ and $\beta^\dagger_{\vec{k}}$ create two different magnon species with energies
\begin{align}
    E_{\vec{k},\alpha} = \varepsilon_{\vec{k}} + \Delta_{\vec{k}}, 
    \quad
    E_{\vec{k},\beta} = \varepsilon_{\vec{k}} - \Delta_{\vec{k}}, 
    \label{eq:magnon-energies}
\end{align}
and
$
    \varepsilon_{\vec{k}} = \sqrt{ A^2_{\vec{k}} - B^2_{\vec{k}} - D^2_{\vec{k}} }
$.
The matrix $T_{\vec{k}}$ obeys $T_{\vec{k}}^\dagger \tau_3 T_{\vec{k}} = \tau_3$ with $\tau_3 = \text{diag}(1,-1)$ \cite{colpa_diagonalization_1978}. It can be written as \cite{Zyuzin2016}
\begin{align}
    T_{\bm{k}} = \begin{pmatrix}
	 	\exp(\mathrm{i} \lambda_{\bm{k}}) \cosh \frac{X_{\bm{k}}}{2} & \sinh \frac{X_{\bm{k}}}{2} \\
		\sinh \frac{X_{\bm{k}}}{2} & \exp(-\mathrm{i}  \lambda_{\bm{k}}) \cosh \frac{X_{\bm{k}}}{2}
  \end{pmatrix},
  \label{eq:T-matrix}
\end{align}
with $\cosh X_{\vec{k}} = A_{\vec{k}} / \varepsilon_{\vec{k}}$, $\sinh X_{\vec{k}} = |\gamma_{\vec{k}}| / \varepsilon_{\vec{k}}$, and $\exp(\mathrm{i} \lambda_{\vec{k}}) = -\gamma_{\vec{k}} / |\gamma_{\vec{k}}|$, where $\gamma_{\vec{k}} = B_{\vec{k}}-\mathrm{i}D_{\vec{k}}$.

Several important observations can be made: 

(1) The block-diagonal structure of $H_{2}$ implies that the $\alpha$ and $\beta$-modes carry opposite spin quantum number $\langle \vec{s} \rangle = \mp \hbar \hat{\vec{N}}$, with the quantization axis given by $\hat{\vec{N}}$. 

(2) According to Eq.~\eqref {eq:magnon-energies}, the magnon band degeneracy is not lifted by DMI but only by the altermagnetic $\Delta_{\vec{k}}$, which enters as a momentum-dependent splitting with $d$-wave symmetry \cite{Smejkal2022altermagnets, SmejkalMagnons23}, as shown along a high-symmetry path [Fig.~\ref{fig:1}(b)] in Fig.~\ref{fig:1}(c). For $\Delta = 0$, the magnons are degenerate over the entire Brillouin zone (see SM \cite{SM}). 

Both observations, (1) and (2), together with the pseudo-Goldstone modes ($E_{\vec{k},\alpha/\beta} \to 0$ as $|\vec{k}|\to 0$), are \emph{artifacts} of the harmonic theory, which we return to later in the context of magnon-magnon interactions. However, these artifacts do not hinder the understanding of the THE to leading order in $1/S$.

(3) More importantly, 
the block diagonalization matrices $T_{\vec{k}}$ in Eq.~\eqref{eq:T-matrix} do not depend on $\Delta_{\vec{k}}$ and are identical for the $\alpha$ and $\beta$ modes. As a result, \emph{the two bands carry identical Berry curvature} given by \cite{Shindou13}
\begin{align}
    \Omega^{\gamma}_{\bm{k},\alpha} = \Omega^{\gamma}_{\bm{k},\beta} = \Omega^{\gamma}_{\bm{k}} =  \mathrm{i} \epsilon_{\alpha \beta \gamma} \left( \tau_{3} \frac{\partial T_{\bm{k}}^{\dagger}}{\partial k_{\alpha}} \tau_{3} \frac{\partial T_{\bm{k}}}{\partial k_{\beta}} \right)_{11}.
\end{align}
This finding highlights the difference to collinear ferromagnets (nondegenerate bands) \cite{Katsura2010, Zhang2013} and antiferromagnets (degenerate bands with opposite Berry curvature) \cite{Cheng2016, Zyuzin2016}. 




%
%

\begin{figure}[]
       \includegraphics[width=\columnwidth]{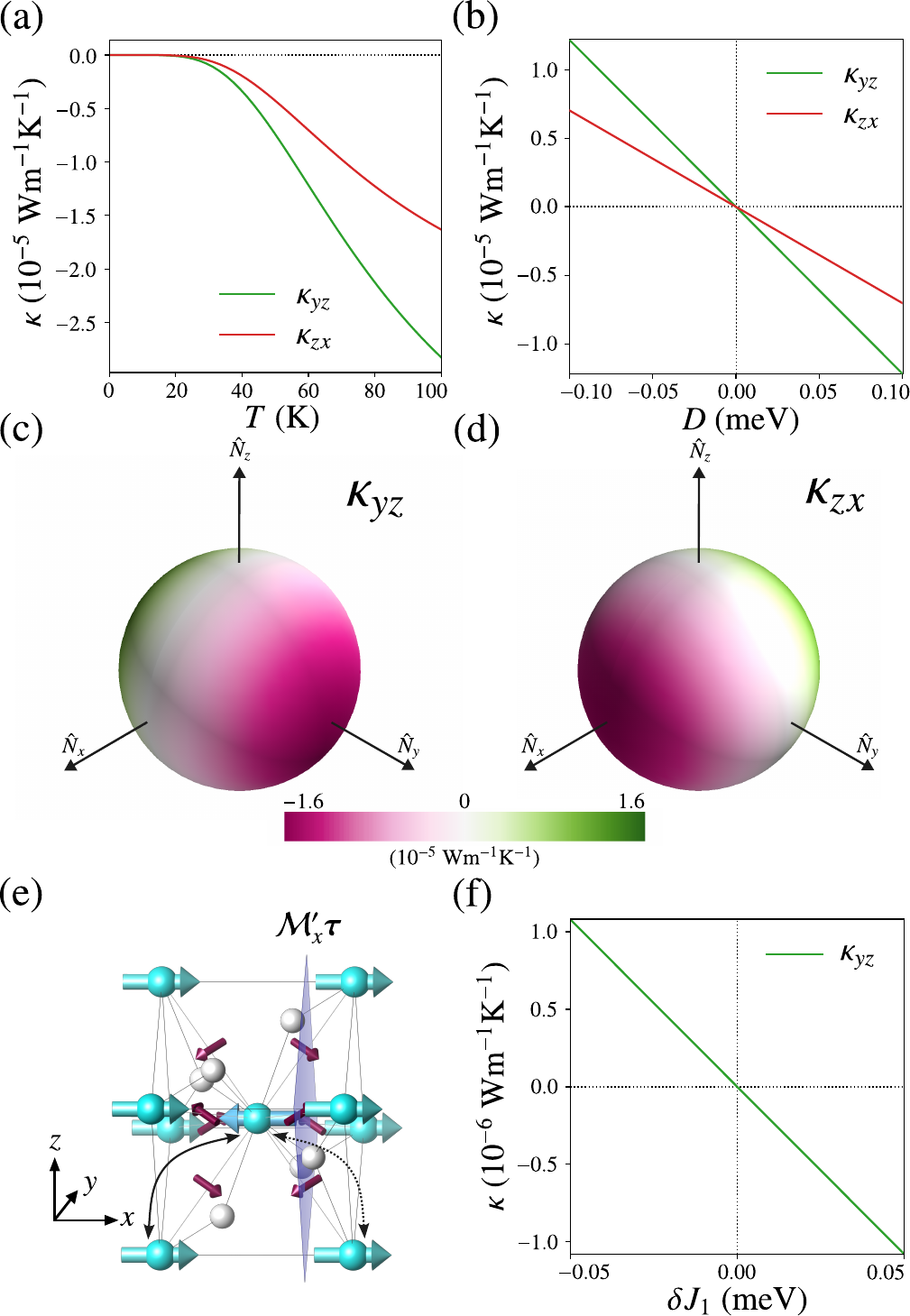}
       \caption{(a) Temperature dependence of the thermal Hall conductivity (THC). (b) Linear dependence of the THC on the Dzyaloshinskii-Moriya interaction $D$. (c), (d) Néel vector dependence of the THC. (e) Antiunitary glide-mirror $\mathcal{M}'_x \tau$ when $\bm{N} \parallel \hat{\bm{x}}$. The solid (dotted) arrow shows strain-renormalized exchanges $J_1 + \delta J_1$ ($J_1 - \delta J_1$). (f) Dependence of the THC on the applied strain, parametrized by $\delta J_1$. We set $\theta = \phi = \pi/3$ for (a)-(b), and the temperature in (b)-(d), (f) is $T = 60$ K. All other parameters as in Fig. \ref{fig:1}.}
       \label{fig:2}
\end{figure}

\paragraph{Thermal Hall effect.}
Using the expression for the intrinsic thermal Hall conductivity derived within a free-boson theory \cite{Matsumoto2011PRL, Matsumoto2011, Matsumoto2014dipolar}, we can express $\vec{\kappa}_\text{H}$ as
\begin{subequations}    
\begin{align}
    \vec{\kappa}_\text{H}
    &=
    -\frac{k_\text{B}^2T}{\hbar V}
    \sum_{\vec{k}}
    \vec{\Omega}_{\vec{k}} \left[
        c_2(\rho(E_{\vec{k},\alpha}))
        +
        c_2(\rho(E_{\vec{k},\beta}))
    \right]
    \label{eq:general-kappa}
    \\
    &\approx
    -\frac{2k_\text{B}^2T}{\hbar V}
    \sum_{\vec{k}}
    \vec{\Omega}_{\vec{k}} 
    c_2(\rho(\varepsilon_{\vec{k}}))
    +
    O(\Delta^2),
    \label{eq:approximate-kappa}
\end{align}
\end{subequations}
where $k_B$ is the Boltzmann constant, $T$ is the temperature, $V$ is the total volume of the system, $\rho(E_{\bm{k}, \alpha/\beta}) = [ \mathrm{e}^{E_{\bm{k}, \alpha/\beta}/(k_B T)} -1 ]^{-1}$ is the Bose-Einstein function, and $c_{2}(x) = \left( 1 + x \right)  \left(\ln{\frac{1 + x}{x}}\right)^2 - \left(\ln{x}\right)^2 - 2 \text{Li}_2(-x)$, where $\text{Li}_2(x)$ is the dilogarithm function.

Since $\vec{\Omega}_{\vec{k}} \propto D$ \cite{SM}, finite $\vec{\kappa}_\text{H}$ requires the relativistic DMI. Importantly, there is no cancellation between bands (in contrast to antiferromagnets \cite{Cheng2016, Zyuzin2016}), and the parity symmetry renders the Berry curvature even in momentum, 
$
    \vec{\Omega}_{\vec{k}} = \vec{\Omega}_{-\vec{k}}
$,
avoiding cancellation between opposite momenta within a single band.
Since $\vec{\Omega}_{\vec{k}}$ does not depend on $\Delta$, a THE can occur even for $\Delta = 0$, as also suggested by the small-$\Delta$ expansion in Eq.~\eqref{eq:approximate-kappa} (see also SM \cite{SM}).
And also vice versa: zero $\vec{\kappa}_\text{H}$ does not allow for conclusions about the size of the altermagnetic spin splitting. Although $\Delta$ and $D$ are ultimately related to the existence of the nonmagnetic ions that break antiferromagnetic symmetries, they arise from independent physical effects: $\Delta$ is a nonrelativistic direction-anisotropic further-neighbor exchange interaction and $D$ is a relativistic nearest-neighbor interaction. Thus, the transport effects they cause---the magnon spin splitter effect~\cite{Naka2019,CuiPRB2023} and the THE, respectively---can arise independently.
Taken together, our observations regarding $\vec{\kappa}_\text{H}$ suggest that there is, in general, no symmetry that would forbid a THE. 

This expectation is verified by a numerical evaluation of Eq.~\eqref{eq:general-kappa}. Figure~\ref{fig:2}(a) shows that $\kappa_{yz}$ and $\kappa_{zx}$ are nonzero for a general orientation of $\vec{N}$, and Fig.~\ref{fig:2}(b) proves their linear dependence on $D$. Figure~\ref{fig:2}(c,d) show $\kappa_{yz}$ and $\kappa_{zx}$ as a function of the N\'{e}el vector orientation. They change sign upon reversing the orientation of $\vec{N}$ (as they should as Hall conductivities), and the intermediate zeros are understood from fine-tuned magnetic point group symmetries:
For example, for $\vec{N} \parallel \hat{\vec{x}}$ ($\vec{N} \parallel \hat{\vec{y}}$) the system holds an anti-unitary glide mirror symmetry $\mathcal{M}'_x \vec{\tau}$ ($\mathcal{M}'_y \tilde{\vec{\tau}}$) with $\vec{\tau} = (0, \frac{1}{2}, \frac{1}{2} )$ [$\tilde{\vec{\tau}} = (\frac{1}{2}, 0, \frac{1}{2} )$], with the mirror in the $yz$ ($xz$) plane, see Fig.~\ref{fig:2}(e), which forbids $\kappa_{yz}$ ($\kappa_{zx}$); see SM for further analysis \cite{SM}.

The intimate relation between $\vec{\kappa}_\text{H}$ and magnetic point group symmetries can be exploited to engineer the THE by strain. Consider the case $\vec{N} \parallel \hat{\vec{x}}$, for which $\kappa_{yz} = 0$. To break the responsible glide mirror symmetry, we apply shear strain along the $[110]$ direction, i.e., we modify the nearest-neighbor exchange couplings $J_1$ so that $J_1 \to J_1 + \delta J_1$ ( $J_1 \to J_1 - \delta J_1$) for bonds with a finite (zero) projection onto $[110]$ \cite{SM}. Indeed, as shown in Fig.~\ref{fig:2}(f), $\kappa_{yz}$ can be activated by strain. 

We recall that as a finite $\vec{\kappa}_\text{H}$ requires magnetic point groups compatible with ferromagnetism, our toy model altermagnet also exhibits a tiny but finite magnetization $\vec{M} = \vec{M}_\text{A} + \vec{M}_\text{B}$. In the SM \cite{SM}, we show that $\vec{M}$ is enabled by thermal fluctuations and comment on fluctuation-induced piezomagnetism. 




%
%

\begin{figure}[]
       \includegraphics[width=\columnwidth]{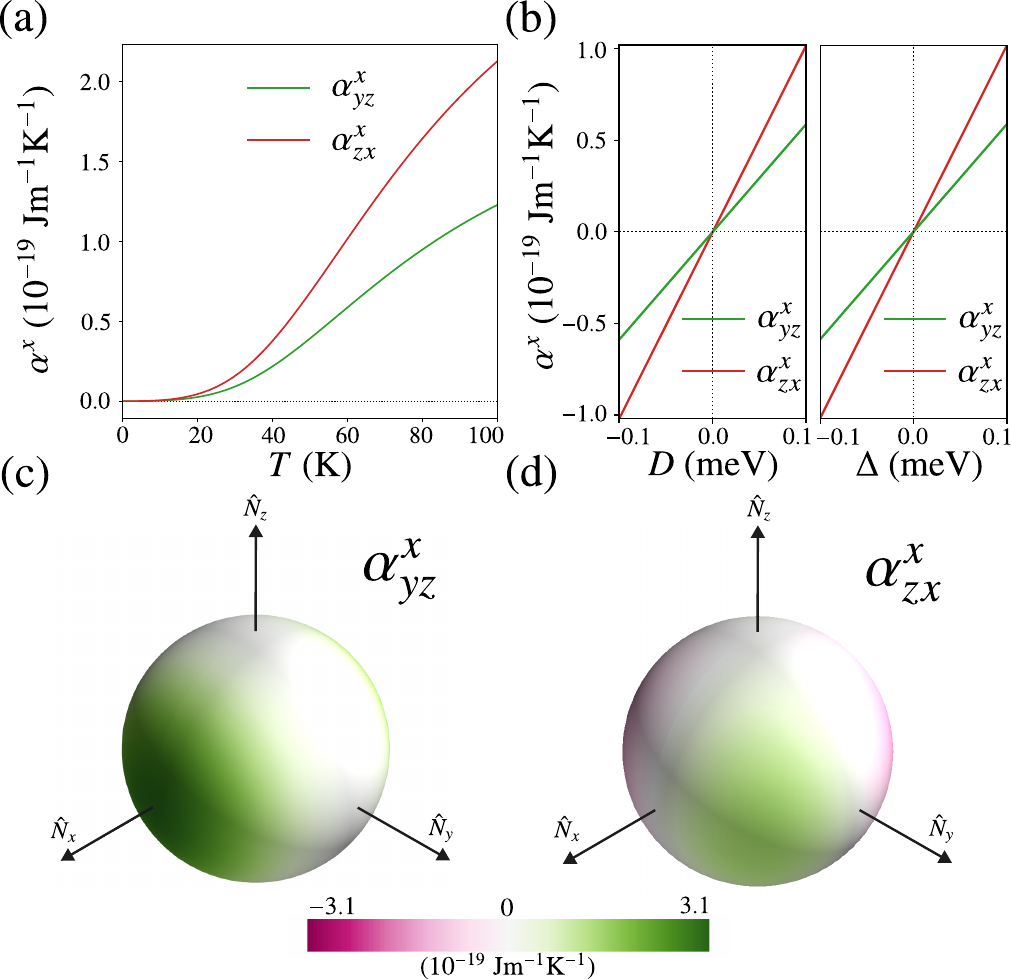}
       \caption{(a) Temperature dependence of the spin Nernst conductivity (SNC). (b) Linear dependence of the SNC on the Dzyaloshinskii-Moriya interaction $D$ (left) and the altermagnetic splitting $\Delta$ (right). (c), (d) Néel vector dependence of the SNC. We set $\theta = \phi = \pi/3$ in (a)-(b) and the temperature in (b)-(d) is $T = 60$ K. All other parameters as in Fig. \ref{fig:1}.}
       \label{fig:3}
\end{figure}

\paragraph{Spin Nernst effect.}
As magnons carry spin angular momentum, a temperature gradient causes a spin current density, $\vec{j}^\gamma = \alpha^\gamma (-\vec{\nabla} T)$. Here, $\alpha^\gamma$ is the thermal spin conductivity and $\gamma$ denotes the transported spin component (parallel to $\bm{N}$). It is known that the altermagnetic splitting causes nonrelativistic spin polarized and spin splitter currents of electrons \cite{Gonzalez2021, Smejkal2022emergent, SmejkalHellenes2022} and magnons \cite{Naka2019, CuiPRB2023}. 
Here, we instead consider the relativistic SNE arising from interband contributions. Relying again on the free-boson transport theory, we write the spin Nernst vector 
$
 \vec{\alpha}^\gamma = (\alpha^\gamma_{yz}, \alpha^\gamma_{zx}, \alpha^\gamma_{xy} )^\text{T}
$ 
as \cite{Cheng2016, Zyuzin2016}
\begin{subequations}    
\begin{align}
    \vec{\alpha}^\gamma
    &=
    -\frac{k_\text{B} \hat{N}^\gamma}{ V}
    \sum_{\vec{k}}
    \vec{\Omega}_{\vec{k}} \left[
    c_1(\rho(E_{\vec{k},\beta}))
    -
    c_1(\rho(E_{\vec{k},\alpha}))
    \right]
    \label{eq:spin-Nernst-exact} \\
    &\approx
    \frac{2 k_\text{B} \hat{N}^\gamma }{ V}
    \sum_{\vec{k}}
    \vec{\Omega}_{\vec{k}} \Delta_{\vec{k}}
    \left.\frac{\partial c_1(\rho(x))}{\partial x}\right|_{x = \varepsilon_{\vec{k}}}
    +
    O(\Delta^3),
    \label{eq:spin-Nernst-approximation}
\end{align}
\end{subequations}
where $c_{1}(x) = \left( 1 + x \right) \ln\left(1 + x\right) - x \ln{x}$.
We evaluate Eq.~\eqref{eq:spin-Nernst-exact} and plot the conductivities $\alpha^{x}_{yz}$ and $\alpha^{x}_{zx}$ as a function of temperature in Fig.~\ref{fig:3}(a), showing a finite SNE. 

In Eq.~\eqref{eq:spin-Nernst-approximation}, we have performed an expansion in small $\Delta$ and read off that a finite $\vec{\alpha}^\gamma$ requires finite $D$ and $\Delta$ \cite{SM}, as also seen numerically in Fig.~\ref{fig:3}(b). This result reflects the following intuitive observation. If $\Delta = 0$, the bands are degenerate, and both magnon species contribute exactly the same to the transverse magnon number current, i.e. the transverse heat current is made in equal parts out of $\alpha$ and $\beta$ magnons. A finite $\Delta$ breaks this degeneracy and establishes an imbalance in transverse $\alpha$ and $\beta$ currents, and, hence, a transverse spin current.

The conductivities are strongly affected by the orientation of $\bm{N}$ [Fig.~\ref{fig:3}(c,d)], which is again a result of magnetic point group symmetries \cite{SM}. Reversing the orientation of $\bm{N}$ does not change the sign of spin Nernst conductivities, because the SNE is a time-reversal even response. 

\paragraph{Magnon-magnon interactions.}
The restriction to the non-interacting theory, $H_2$, leads to artificial spin conservation and pseudo-Goldstone modes [recall observations (1) and (2)], and to a vanishing $\kappa_{xy}$ and $\alpha^{\gamma}_{xy}$, as we explain in \cite{SM}. These observations are the result of spurious symmetries \cite{Rau2018PseudoGoldstone,Gohlke2023PRL} that arise because $H_2$ only contains terms $\propto \vec{\hat{N}} \cdot \hat{\vec{D}}_i$ [see Eq.~\eqref{eq:DMIinH2}] but the components of the DMI vectors perpendicular to $\vec{\hat{N}}$ drop out. Their symmetry-breaking effect is hence only seen by 
the three-magnon interactions,
\begin{align}
    H_3 = \sum_{\bm{k}, \bm{q}} \left( D_{\bm{k}}^- a_{\bm{k}}^\dagger b_{\bm{q}}^\dagger b_{\bm{k} + \bm{q}} + D_{\bm{k}}^+ b_{\bm{k}}^\dagger a_{\bm{q}}^\dagger a_{\bm{k} + \bm{q}} + \mathrm{H.c.}\right),
\end{align}
where 
$D_{\bm{k}}^{\pm} \propto D$ is defined in \cite{SM}.
After a Bogoliubov transformation,
$H_3$ contains terms such as
$\alpha^\dagger \alpha^\dagger \beta^\dagger$ and
$\alpha^\dagger \alpha \beta$ that
explicitly break spin conservation and hybridize the two magnon species, thereby 
lifting the spurious symmetry and giving rise to finite $\kappa_{xy}$ and $\alpha^{\gamma}_{xy}$ (see SM \cite{SM} for further details).


%
%

%
%

\begin{table}[]
  \centering
  \begin{tabular}{  b{1cm}   c  c  c  }
    \toprule
     & Ferromagnet & Antiferromagnet & Altermagnet \\ 
     & \begin{minipage}{.25\linewidth}
      \includegraphics[width=\linewidth]{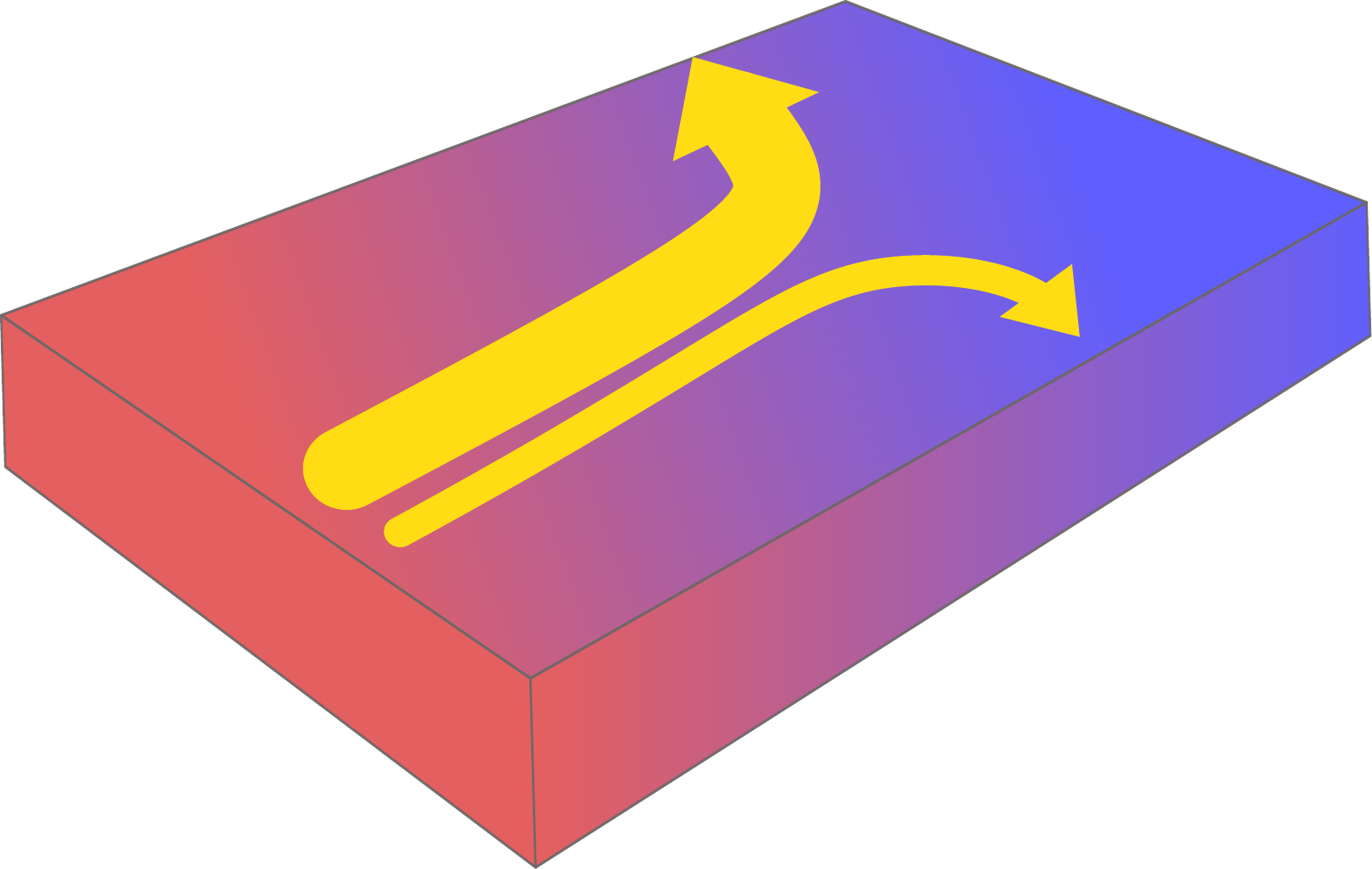}
    \end{minipage} &
    \begin{minipage}{.25\linewidth}
      \includegraphics[width=\linewidth]{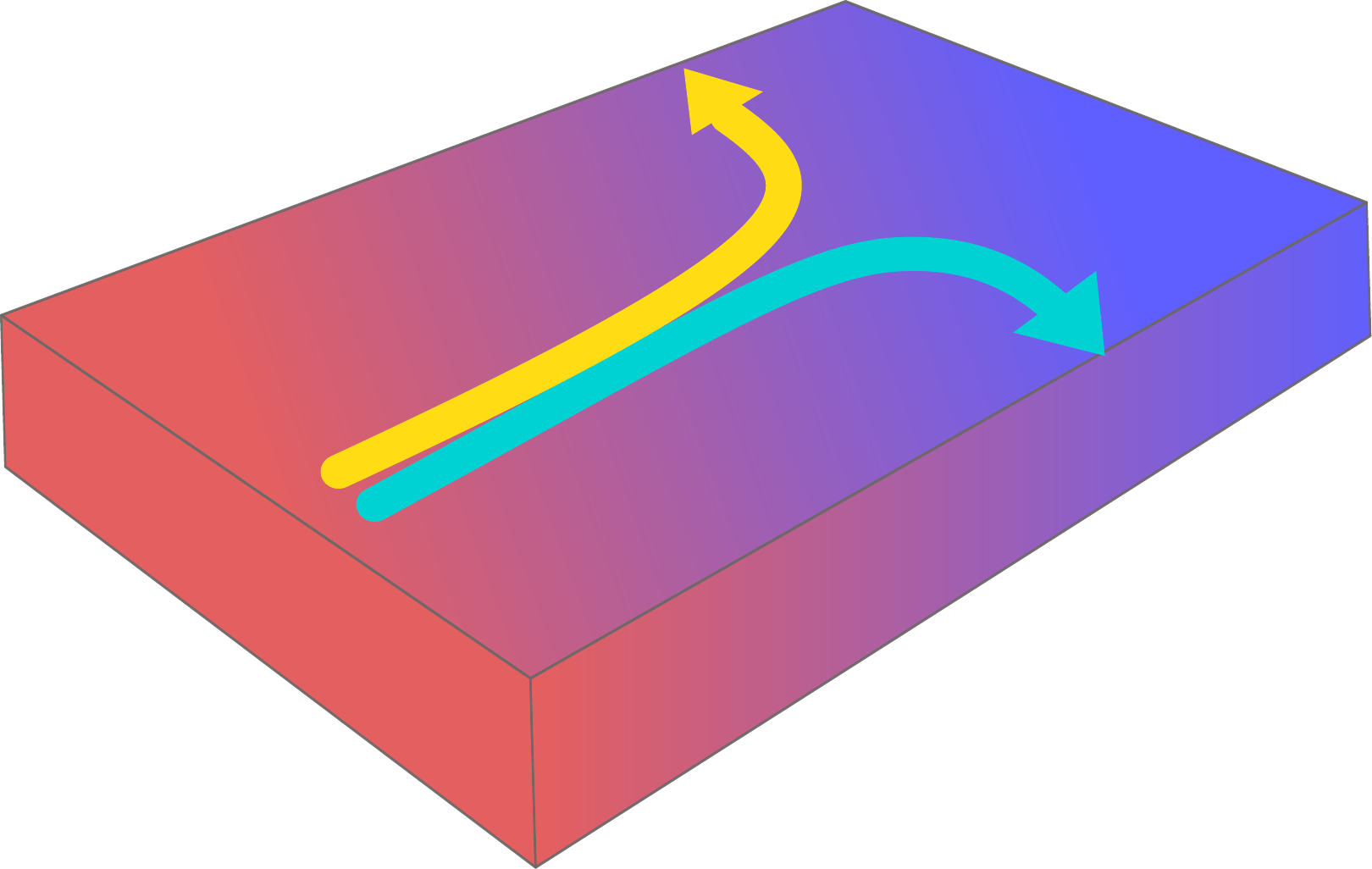}
    \end{minipage} & 
    \begin{minipage}{.25\linewidth}
      \includegraphics[width=\linewidth]{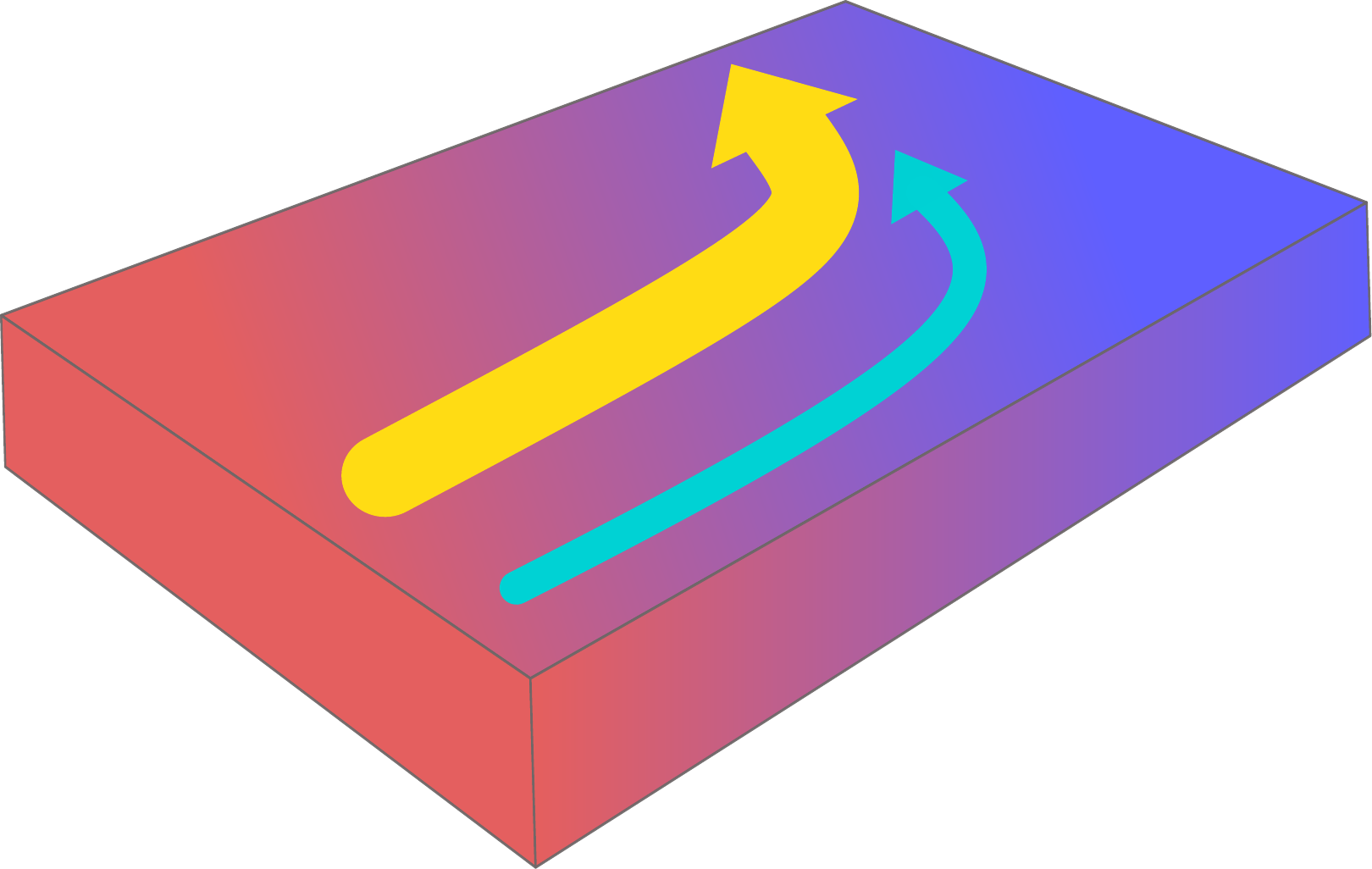}
    \end{minipage} \vspace{0.1cm}\\ \hline 
   
    THE & \cmark
    &
    \xmark
    & 
     \cmark \\ \hline
    SNE& \cmark
    &
     \cmark
    & 
     \cmark
    \\ \botrule 
    \end{tabular}
    \caption{
    Comparison of DMI-induced THE and SNE in spin-conserving two-band ferromagnets, antiferromagnets, and altermagnets. Heat flows from red (high temperature) to blue (low temperature). The cyan (yellow) arrows represent magnon currents with spin up (down). The thickness of the arrows reflects the magnitude of the transverse currents.
    }
    \label{table:1}
\end{table}


\paragraph{Discussion and Conclusion.}
We have presented a minimal model of an insulating ($d$-wave) altermagnet exhibiting a crystal THE with a conductivity $\bm{\kappa}_{\text{H}}$ that is highly tunable by the N\'{e}el vector orientation and strain.
Taken together with our results on the SNE, we summarize in Tab.~\ref{table:1} how the 
DMI-induced magnonic transverse heat and spin currents in altermagnets differ from those in collinear ferromagnets and antiferromagnets. (We consider systems with two magnon bands and spin conservation at the level of the harmonic theory.) In ferromagnets, the nondegenerate magnon bands with opposite Berry curvature and unequal thermal population cause net transverse heat and spin currents. In antiferromagnets, the magnon bands are spin-degenerate and their Berry curvature is opposite, leading to zero heat current (no THE) but to a finite SNE. In altermagnets, the magnon bands carry identical Berry curvature, causing a transverse heat current, accompanied by a spin current only for finite altermagnetic band splitting.
In contrast to the nonrelativistic magnon spin-splitter current \cite{Naka2019, CuiPRB2023}, the relativistic transport effects discussed here are expected to exist not only in $d$-wave altermagnets, but also, for example, in $g$-wave altermagnets like hematite.

Our model resembles the insulating magnetic rutiles MnF$_2$, NiF$_2$, and CoF$_2$, for which an altermagnetic magnon splitting has not been reported in experiments. Indeed, our \textit{ab initio} calculations for CoF$_2$ predict only a tiny splitting (see SM \cite{SM}), which, however, as established above, still can lead to a sizable and measurable THE. We therefore suggest to explore the THE, for example, in the easy-plane magnet NiF$_2$, for which further theoretical modelling should include the local magnetocrystalline anisotropies \cite{Moriya1960NiF2}.

\begin{acknowledgments}
\paragraph{Acknowledgments.}
This work was funded by the Deutsche Forschungsgemeinschaft (DFG, German Research Foundation) -- Project No.~504261060 (Emmy Noether Programme). L.S. acknowledges support from the Johannes Gutenberg-Universität Mainz TopDyn initiative.
\end{acknowledgments}

\bibliography{bib}

\end{document}


\title{Supplementary Material to: Spontaneous Crystal Thermal Hall Effect in Insulating Altermagnets}%

\author{Rhea Hoyer}
\affiliation{Department of Physics, Johannes Gutenberg University Mainz, 55128 Mainz, Germany}

\author{Rodrigo Jaeschke-Ubiergo}
\affiliation{Department of Physics, Johannes Gutenberg University Mainz, 55128 Mainz, Germany}

\author{Kyo-Hoon Ahn}
\affiliation{Institute of Physics, Czech Academy of Sciences, Cukrovarnická 10, 162 00 Praha 6, Czech Republic}

\author{Libor Šmejkal}
\affiliation{Department of Physics, Johannes Gutenberg University Mainz, 55128 Mainz, Germany}
\affiliation{Institute of Physics, Czech Academy of Sciences, Cukrovarnická 10, 162 00 Praha 6, Czech Republic}

\author{Alexander Mook}
\affiliation{Department of Physics, Johannes Gutenberg University Mainz, 55128 Mainz, Germany}

\begin{abstract}
    In this Supplemental Material we provide additional information on (S.I) the (nonlinear) spin-wave theory, the derivation of the magnon dispersion, and the magnon Berry curvature, (S.II) the calculation of the thermal Hall conductivity, (S.III) symmetries of the thermal conductivity tensor, (S.IV) the implementation of strain, (S.V) the magnetization induced by thermal fluctuations and piezomagnetism, (S.VI) the calculation of the spin Nernst conductivity and the symmetries of the related conductivity tensor, (S.VII) spurious symmetries of the harmonic theory and how they get lifted by magnon-magnon interactions, and (S.VIII) proposed material candidates.
\end{abstract}

\date{\today}%
\maketitle

\tableofcontents

\section{Spin model and magnon spectrum}
The effective spin Hamiltonian written term by term reads 
\begin{align}
	H =& \,J_{1} \sum_{\langle i, j \rangle} \bm{S}_{i}^{\text{A}/\text{B}} \cdot \bm{S}_{j}^{\text{B}/\text{A}} + J_{2} \sum_{\langle \langle i, j \rangle \rangle} \bm{S}_{i}^{\text{A}/\text{B}} \cdot \bm{S}_{j}^{\text{A}/\text{B}} + J_{3} \sum_{\langle \langle \langle i, j \rangle \rangle \rangle} \bm{S}_{i}^{\text{A}/\text{B}} \cdot \bm{S}_{j}^{\text{A}/\text{B}} + \Delta \left( \sum_{\langle \langle \langle i, j \rangle \rangle \rangle^{xy}} \bm{S}_{i}^{\text{A}} \cdot \bm{S}_{j}^{\text{A}} -  \sum_{\langle \langle \langle i, j \rangle \rangle \rangle^{x\overline{y}}} \bm{S}_{i}^{\text{A}} \cdot \bm{S}_{j}^{\text{A}} \right)\nonumber \\
	& - \Delta \left( \sum_{\langle \langle \langle i, j \rangle \rangle \rangle^{xy}} \bm{S}_{i}^{\text{B}} \cdot \bm{S}_{j}^{\text{B}} -  \sum_{\langle \langle \langle i, j \rangle \rangle \rangle^{x\overline{y}}} \bm{S}_{i}^{\text{B}} \cdot \bm{S}_{j}^{\text{B}} \right) 
	 + \sum_{\langle i, j \rangle } \bm{D}_{ij} \cdot  \left( \bm{S}_{i}^{\text{A}/\text{B}} \times \bm{S}_{j}^{\text{B}/\text{A}} \right),
\end{align}
where $J_1$, $J_2$, and $J_3$ denote first-, second-, and third-nearest neighbor exchange interactions. The terms proportional to the altermagnetic directional exchange anisotropy $\Delta$ are defined for third-nearest neighbors, where the superscript in $\langle \langle \langle i, j \rangle \rangle \rangle^{xy}$ defines the bond direction. The Dzyaloshinskii-Moriya interactions (DMI) with vectors $\bm{D}_{ij}$, which are defined in the main text, is restricted to nearest neighbors.

\subsection{Spin-to-boson transformation}
We define the orthonormal basis $\{\hat{\bm{x}}, \hat{\bm{y}}, \hat{\bm{z}}\}$ for both the A and B sublattice with 
\begin{subequations}
\begin{align}
    \hat{\bm{x}} & = \left(\cos \theta \cos \phi, \cos \theta \sin \phi, \sin \theta \right)^{\mathrm{T}}, \\
    \hat{\bm{y}} & = \left(- \sin \phi, \cos \phi, 0 \right)^{\mathrm{T}}, \\
    \hat{\bm{z}}  = \hat{\bm{N}} & = \left(\sin \theta \cos \phi, \sin \theta \sin \phi, \cos \theta \right)^{\mathrm{T}}, \label{eq:N}
\end{align}
\label{eq:coordinate-system}
\end{subequations}
and write the spins as
\begin{align}
    \bm{S}_i^{\text{A}/\text{B}} = \widetilde{S}^{\text{A}/\text{B}, x}_i \hat{\bm{x}} +  \widetilde{S}^{\text{A}/\text{B}, y}_i \hat{\bm{y}} +  \widetilde{S}^{\text{A}/\text{B}, z}_i \hat{\bm{z}}.
\end{align}
We define the Holstein-Primakoff transformation from spins to bosons \cite{holsteinprimakoff1940} up to second order in the square root and assume a collinear ground state, such that
\begin{subequations}
    \begin{align}
        \widetilde{S}^{\text{A}, x}_i & \approx \frac{\sqrt{2S}}{2} \left( a_i + a_i^\dagger - \frac{a_i^\dagger a_i  a_i}{4 S}  - \frac{a_i^\dagger a_i^\dagger  a_i}{4 S} \right), \\
        \widetilde{S}^{\text{A}, y}_i & \approx  \frac{\sqrt{2S}}{2 \mathrm{i}} \left( a_i - a_i^\dagger - \frac{a_i^\dagger a_i  a_i}{4 S}  + \frac{a_i^\dagger a_i^\dagger  a_i}{4 S} \right), \\
        \widetilde{S}^{\text{A}, z}_i &= S - a^\dagger_i a_i, \\
        \widetilde{S}^{\text{B}, x}_i &\approx \frac{\sqrt{2S}}{2} \left( b_i + b_i^\dagger - \frac{b_i^\dagger b_i  b_i}{4 S}  - \frac{b_i^\dagger b_i^\dagger b_i}{4 S} \right), \\
        \widetilde{S}^{\text{B}, y}_i &\approx  -\frac{\sqrt{2S}}{2 \mathrm{i}} \left( b_i - b_i^\dagger - \frac{b_i^\dagger b_i  b_i}{4 S}  + \frac{b_i^\dagger b_i^\dagger  b_i}{4 S} \right), \\
        \widetilde{S}^{\text{B}, z}_i &= - S + b^\dagger_i b_i.
    \end{align}
\end{subequations}
The $a_i^\dagger$ ($b_i^\dagger$) and $a_i$ ($b_i$) correspond to the bosonic creation and annihilation operators of sublattice A (B), respectively, with the usual commutation relations $[a_i, a_j^\dagger] = [b_i, b_j^\dagger] = \delta_{ij}$.

We then expand the Hamiltonian in terms of numbers of bosons,
\begin{align}
    H = H_0 + H_1 +H_2 + H_3 + H_4 + \ldots \quad .
\end{align}
We find that the contributions from the Heisenberg exchanges only lead to $H_0$, $H_2$, $H_4$ and higher terms with an even number of bosons. We Fourier transform the bosonic operators, 
\begin{align}
    a_{i}^\dagger/b_{i}^\dagger = \frac{1}{\sqrt{N}} \sum_{\bm{k}} \mathrm{e}^{-\mathrm{i} \bm{k} \cdot \bm{r}_{i}} a_{\bm{k}}^\dagger/b_{\bm{k}}^\dagger, \qquad a_{i}/b_{i} = \frac{1}{\sqrt{N}} \sum_{\bm{k}} \mathrm{e}^{\mathrm{i} \bm{k} \cdot \bm{r}_{i}} a_{\bm{k}}/b_{\bm{k}},
    \label{eq:FT}
\end{align}
and the bilinear $H_2 = H^{J_ 1}_2 + H^{J_ 2}_2 + H^{J_ 3}_2 + H^{\Delta}_2$ terms thus read
\begin{subequations}
    \begin{align}
        H^{J_ 1}_2 & = 2 J_ 1 S \sum_{\bm{k}} \left[ 4 \left( a_ {\bm{k}}^ \dagger a_{\bm{k}}  + b_{\bm{k}}^\dagger b_{\bm{k}} \right) + \sum_{j = 1}^4 \cos ( \bm{k} \cdot \bm{\delta}_j ) \left( a_{\bm{k}} b_{-\bm{k}} + a_ {\bm{k}}^\dagger b_ {-\bm{k}}^\dagger \right) \right], \\
        H^{J_ 2}_2 &= 2 J_ 2 S \sum_{\bm{k}} \left( \cos k_x + \cos k_y + \cos k_z - 3\right) \left( a_ {\bm{k}}^\dagger a_{\bm{k}}  + b_ {\bm{k}}^\dagger b_{\bm{k}} \right), \\
        H^{J_ 3}_2 &= 2 J_ 3 S \sum_{\bm{k}} \left[ \cos (k_x + k_y ) + \cos (k_x - k_y) -2 \right] \left( a_{\bm{k}}^\dagger a_{\bm{k}}  + b_{\bm{k}}^\dagger b_{\bm{k}} \right), \\
        H^{\Delta}_2 & = 2 \Delta S \sum_{\bm{k}} \left[ \cos (k_x + k_y ) - \cos (k_x - k_y) \right] \left( a_{\bm{k}}^\dagger a_{\bm{k}}  -b_{\bm{k}}^\dagger b_{\bm{k}} \right),
    \end{align}
\end{subequations}
where we define the $\bm{\delta}_j$'s as
\begin{align}
        \bm{\delta}_{1} = \frac{1}{2}\left(1, 1, 1\right)^\text{T}, \quad \bm{\delta}_{2} = \frac{1}{2}\left(1, 1, -1\right)^\text{T}, \quad \bm{\delta}_{3} = \frac{1}{2}\left(1, -1, 1\right)^\text{T} , \quad \text{and} \quad \bm{\delta}_{4} = \frac{1}{2}\left(-1, 1, 1\right)^\text{T},
\end{align}
which are the nearest-neighbor links. To lighten notation, we will from now on omit the index $j$ of these links.

Next, we derive the bosonic representation of the DMI, which we can write as
\begin{subequations}
\begin{align}
    H_{\text{DMI}}
    =& \, \frac{1}{2} \sum_{i \in \text{A}} \sum_{\delta} \bm{D}_{\delta} \cdot \left( \hat{\bm{N}} \widetilde{S}_i^{\text{A}, x} \widetilde{S}_{i + \delta}^{\text{B}, y}  - \hat{\bm{y}} \widetilde{S}_i^{\text{A}, x} \widetilde{S}_{i + \delta}^{\text{B}, z}  - \hat{\bm{N}} \widetilde{S}_i^{\text{A}, y} \widetilde{S}_{i + \delta}^{\text{B}, x} + \hat{\bm{x}} \widetilde{S}_i^{\text{A}, y} \widetilde{S}_{i + \delta}^{\text{B}, z} + \hat{\bm{y}} \widetilde{S}_i^{\text{A}, z} \widetilde{S}_{i + \delta}^{\text{B}, x} - \hat{\bm{x}} \widetilde{S}_i^{\text{A}, z} \widetilde{S}_{i + \delta}^{\text{B}, y} \right)  \nonumber \\
    & - \frac{1}{2} \sum_{i \in \text{B}} \sum_{\delta} \bm{D}_{\delta} \cdot \left( \hat{\bm{N}} \widetilde{S}_i^{\text{B}, x} \widetilde{S}_{i + \delta}^{\text{A}, y}  - \hat{\bm{y}} \widetilde{S}_i^{\text{B}, x} \widetilde{S}_{i + \delta}^{\text{A}, z}  - \hat{\bm{N}} \widetilde{S}_i^{\text{B}, y} \widetilde{S}_{i + \delta}^{\text{A}, x} + \hat{\bm{x}} \widetilde{S}_i^{\text{B}, y} \widetilde{S}_{i + \delta}^{\text{A}, z} + \hat{\bm{y}} \widetilde{S}_i^{\text{B}, z} \widetilde{S}_{i + \delta}^{\text{A}, x} - \hat{\bm{x}} \widetilde{S}_i^{\text{B}, z} \widetilde{S}_{i + \delta}^{\text{A}, y} \right) \label{eq:DMI-z}\\ 
     =& \,H_{D}^{\perp \hat{\bm{N}}} + H_{D}^{\parallel \hat{\bm{N}}} ,  \label{eq:DMIparaperp}
\end{align}
\end{subequations}
where in Eq.~\eqref{eq:DMI-z} we used $\hat{\bm{z}} = \hat{\bm{N}}$ [cf. Eqs.~\eqref{eq:coordinate-system}].
The part $H_{D}^{\perp \hat{\bm{N}}}$, which consists of terms in which $\bm{D}_{\delta}$ gets dotted into $\hat{\bm{x}}$ and $\hat{\bm{y}}$, leads to an odd number of bosons, that is, to $H_1$, $H_3$ and higher odd-boson terms. To see so, we consider
\begin{subequations}
    \begin{align}
        H_{D}^{\perp \hat{\bm{N}}} = &\, \frac{1}{2} \sum_{i \in \text{A}} \sum_{\delta} \bm{D}_{\delta} \cdot \left( - \hat{\bm{y}} \widetilde{S}_i^{\text{A}, x} \widetilde{S}_{i + \delta}^{\text{B}, z}  + \hat{\bm{x}} \widetilde{S}_i^{\text{A}, y} \widetilde{S}_{i + \delta}^{\text{B}, z} + \hat{\bm{y}} \widetilde{S}_i^{\text{A}, z} \widetilde{S}_{i + \delta}^{\text{B}, x} - \hat{\bm{x}} \widetilde{S}_i^{\text{A}, z} \widetilde{S}_{i + \delta}^{\text{B}, y} \right)  \nonumber \\
    & - \frac{1}{2} \sum_{i \in \text{B}} \sum_{\delta} \bm{D}_{\delta} \cdot \left( - \hat{\bm{y}} \widetilde{S}_i^{\text{B}, x} \widetilde{S}_{i + \delta}^{\text{A}, z} + \hat{\bm{x}} \widetilde{S}_i^{\text{B}, y} \widetilde{S}_{i + \delta}^{\text{A}, z} + \hat{\bm{y}} \widetilde{S}_i^{\text{B}, z} \widetilde{S}_{i + \delta}^{\text{A}, x} - \hat{\bm{x}} \widetilde{S}_i^{\text{B}, z} \widetilde{S}_{i + \delta}^{\text{A}, y} \right), 
    \\
    = &\, H_1^D + H_3^D + \ldots, \label{eq:H1-H3}
    \end{align}
\end{subequations}
where
\begin{subequations}
    \begin{align}
         H_1^D =&\, \frac{S}{2} \sum_{i \in \text{A}} \sum_{\delta} \bm{D}_{\delta} \cdot \left( \hat{\bm{y}} \widetilde{S}_i^{\text{A}, x}   - \hat{\bm{x}} \widetilde{S}_i^{\text{A}, y}  + \hat{\bm{y}}  \widetilde{S}_{i + \delta}^{\text{B}, x} - \hat{\bm{x}} \widetilde{S}_{i + \delta}^{\text{B}, y} \right)  - \frac{S}{2} \sum_{i \in \text{B}} \sum_{\delta} \bm{D}_{\delta} \cdot \left( - \hat{\bm{y}} \widetilde{S}_i^{\text{B}, x}  + \hat{\bm{x}} \widetilde{S}_i^{\text{B}, y}  - \hat{\bm{y}} \widetilde{S}_{i + \delta}^{\text{A}, x} + \hat{\bm{x}} \widetilde{S}_{i + \delta}^{\text{A}, y} \right) \\
         =&\, \frac{S \sqrt{2S}}{4} \sum_{i \in \text{A}} \sum_{\delta} \bm{D}_{\delta} \cdot \left[\hat{\bm{y}} \left(  a_i + a_i^\dagger \right)   + \mathrm{i} \hat{\bm{x}} \left(  a_i - a_i^\dagger \right)  + \hat{\bm{y}}  \left(  b_{i + \delta}^\dagger +b_{i + \delta} \right) + \mathrm{i}  \hat{\bm{x}} \left(  b_{i + \delta}^\dagger - b_{i + \delta} \right) \right]  \nonumber \\
         & + \frac{S \sqrt{2S}}{4} \sum_{i \in \text{B}} \sum_{\delta} \bm{D}_{\delta} \cdot \left[ \hat{\bm{y}} \left(  b_i^\dagger + b_i \right)   + \mathrm{i} \hat{\bm{x}}\left(  b_i^\dagger - b_i \right)  + \hat{\bm{y}} \left( a_{i + \delta} +a_{i + \delta}^\dagger \right) + \mathrm{i} \hat{\bm{x}} \left( a_{i + \delta} -a_{i + \delta}^\dagger \right) \right]
    \end{align}
\end{subequations}
and $H_3^D$ will be given in Sec.~\ref{sec:mminteractions} in the context of magnon-magnon interactions. 
Because of $\sum_{\delta} \bm{D}_{\delta}  = 0 $, we can write 
\begin{align}
     H_1^D =&\,\frac{S \sqrt{2S}}{4} \sum_{i \in \text{A}} \sum_{\delta} \bm{D}_{\delta} \cdot \left[\hat{\bm{y}}  \left(  b_{i + \delta}^\dagger +b_{i + \delta} \right) + \mathrm{i}  \hat{\bm{x}} \left(  b_{i + \delta}^\dagger - b_{i + \delta} \right) \right]   + \frac{S \sqrt{2S}}{4} \sum_{i \in \text{B}} \sum_{\delta} \bm{D}_{\delta} \cdot \left[  \hat{\bm{y}} \left( a_{i + \delta} +a_{i + \delta}^\dagger \right) + \mathrm{i} \hat{\bm{x}} \left( a_{i + \delta} -a_{i + \delta}^\dagger \right) \right],
\end{align}
and the $i + \delta$ terms vanish for the same reason, leading to $ H_1^D = 0$. 
Thus, the DMI does not cause a classical weak ferromagnetic moment and the ground state order is collinear.

We now turn to $H_{D}^{\parallel \hat{\bm{N}}}$ in Eq.~\eqref{eq:DMIparaperp}, which contains terms in which $\bm{D}_\delta$ gets dotted into $\hat{\bm{N}}$,
\begin{align}
        H_{D}^{\parallel \hat{\bm{N}}} = &\, \frac{1}{2} \sum_{i \in \text{A}} \sum_{\delta} \bm{D}_{\delta} \cdot \hat{\bm{N}} \left(  \widetilde{S}_i^{\text{A}, x} \widetilde{S}_{i + \delta}^{\text{B}, y}  -  \widetilde{S}_i^{\text{A}, y} \widetilde{S}_{i + \delta}^{\text{B}, x} \right) -  \frac{1}{2} \sum_{i \in \text{B}} \sum_{\delta} \bm{D}_{\delta} \cdot \hat{\bm{N}} \left(  \widetilde{S}_i^{\text{B}, x} \widetilde{S}_{i + \delta}^{\text{A}, y}  -  \widetilde{S}_i^{\text{B}, y} \widetilde{S}_{i + \delta}^{\text{A}, x} \right).
\end{align}
After a magnon expansion, we obtain terms with an even number of bosons, $H_{D}^{\parallel \hat{\bm{N}}} = H_2^D + H_4^D + \ldots$; in particular, we find the bilinear contribution
\begin{subequations}
    \begin{align}
        H_2^D
        = &\, -\frac{\mathrm{i} S}{4} \sum_{i \in \text{A}} \sum_{\delta} \hat{\bm{N}} \cdot \bm{D}_{\delta} \left( a_i b_{i + \delta}^\dagger - a_i b_{i + \delta} + a_i^\dagger b_{i + \delta}^\dagger - a_i^\dagger b_{i + \delta} - a_i b_{i + \delta}^\dagger - a_i b_{i + \delta} + a_i^\dagger b_{i + \delta}^\dagger + a_i^\dagger b_{i + \delta}\right) \nonumber \\
        & + \frac{\mathrm{i} S}{4} \sum_{i \in \text{B}} \sum_{\delta} \hat{\bm{N}} \cdot \bm{D}_{\delta} \left( b_i^\dagger a_{i + \delta} - b_i^\dagger a_{i + \delta}^\dagger + b_i a_{i + \delta} - b_i a_{i + \delta}^\dagger - b_i^\dagger a_{i + \delta} - b_i^\dagger a_{i + \delta}^\dagger + b_i a_{i + \delta} + b_i a_{i + \delta}^\dagger\right) \\
    = &\, \mathrm{i} S \sum_{\bm{k}} \sum_{\delta} \hat{\bm{N}} \cdot \bm{D}_{\delta} \cos (\bm{k} \cdot \bm{\delta} ) \left( a_{\bm{k}} b_{-\bm{k}} +  a_{-\bm{k}} b_{\bm{k}} -  a_{\bm{k}}^\dagger b_{-\bm{k}}^\dagger - a_{-\bm{k}}^\dagger b_{\bm{k}}^\dagger\right) \\
        = &\, \mathrm{i} 2 D S \sum_{\bm{k}} \sum_{\delta} \hat{\bm{N}} \cdot \hat{\bm{D}}_{\delta} \cos (\bm{k} \cdot \bm{\delta} ) \left( a_{\bm{k}} b_{-\bm{k}} -  a_{\bm{k}}^\dagger b_{-\bm{k}}^\dagger \right).
    \end{align}
\end{subequations}

\subsection{Noninteracting theory of magnons}
In total, the resulting $\bm{k}$-dependent harmonic Hamiltonian reads 
\begin{align}
	H_2 =&  \sum_{\bm{k}} A_{\bm{k}}  \left(a_{\bm{k}}^{\dagger} a_{\bm{k}} + b_{\bm{k}}^{\dagger} b_{\bm{k}} \right) + \Delta_{\bm{k}}  \left(a_{\bm{k}}^{\dagger} a_{\bm{k}} - b_{\bm{k}}^{\dagger} b_{\bm{k}} \right)  +  B_{\bm{k}} \left( a_{\bm{k}} b_{-k} + a_{\bm{k}}^{\dagger} b_{-\bm{k}}^{\dagger}  \right) + \mathrm{i} D_{\bm{k}} \left( a_{\bm{k}} b_{-k}  - a_{\bm{k}}^{\dagger} b_{-\bm{k}}^{\dagger} \right)  \\ 
	=& \frac{1}{2} \sum_{\bm{k}} \sum_{\sigma=\pm} \bm{\Psi}_{\bm{k}, \sigma}^{\dagger} H_{\bm{k}, \sigma} \bm{\Psi}_{\bm{k}, \sigma},
\end{align}
with $ \bm{\Psi}_{\bm{k}, +}^{\dagger} = \left(a_{\bm{k}}^{\dagger}, b_{-\bm{k}}\right)$ and $\bm{\Psi}_{\bm{k}, -}^{\dagger} = \left(b_{\bm{k}}^{\dagger}, a_{-\bm{k}}\right),$ and 
\begin{align}
\label{sm:eq:hamilton-matrix}
	H_{\bm{k}, \pm} = \begin{pNiceMatrix}
		A_{\bm{k}} \pm \Delta_{\bm{k}} & B_{\bm{k}} - \mathrm{i} D_{\bm{k}} \\
		B_{\bm{k}} + \mathrm{i} D_{\bm{k}} & A_{\bm{k}} \mp \Delta_{\bm{k}} 
	\end{pNiceMatrix}.
\end{align}
The components of the Hamilton kernels~\eqref{sm:eq:hamilton-matrix} read
\begin{align}
	A_{\bm{k}} = \,&  S\left\{8 J_{1} - 6  J_{2}  - 4 J_{3}  + 2 J_{2} \left(\cos k_{x} +\cos k_{y} + \cos k_{z} \right)  + 2 J_{3} \left[\cos (k_{x} + k_{y}) + \cos (k_{x} - k_{y}) \right] \right\}, \\
	\Delta_{\bm{k}} =\,&  2 \Delta S \left[\cos (k_{x} + k_{y}) - \cos (k_{x} - k_{y}) \right], \\
	B_{\bm{k}} =\,&  2 J_{1} S \left( \cos \frac{k_{x} + k_{y} + k_{z}}{2} + \cos \frac{k_{x} + k_{y} - k_{z}}{2} + \cos \frac{k_{x} - k_{y} + k_{z}}{2} + \cos \frac{-k_{x} + k_{y} + k_{z}}{2} \right), \\
	D_{\bm{k}} =\,&  2 D S \hat{\bm{N}} \cdot \left( \hat{\bm{D}}_{1} \cos \frac{k_{x} + k_{y} + k_{z}}{2} + \hat{\bm{D}}_{2}\cos \frac{k_{x} + k_{y} - k_{z}}{2} + \hat{\bm{D}}_{3} \cos \frac{k_{x} - k_{y} + k_{z}}{2} + \hat{\bm{D}}_{4}\cos \frac{-k_{x} + k_{y} + k_{z}}{2}\right), 
\end{align}
with 
\begin{align}
 	\hat{\bm{D}}_{1} = \frac{1}{\sqrt{2}} \begin{pNiceMatrix}
	 1 \\
	 -1 \\ 
	  0\\
	\end{pNiceMatrix}, \qquad
	\hat{\bm{D}}_{2} = \frac{1}{\sqrt{2}} \begin{pNiceMatrix}
	 -1 \\
	 1 \\ 
	 0 \\
	\end{pNiceMatrix}, \qquad
	\hat{\bm{D}}_{3} = \frac{1}{\sqrt{2}} \begin{pNiceMatrix}
	 1 \\
	 1 \\ 
	 0 \\
	\end{pNiceMatrix}, \qquad
	\hat{\bm{D}}_{4} = \frac{1}{\sqrt{2}} \begin{pNiceMatrix}
	 -1 \\
	 -1 \\ 
	 0 \\
	\end{pNiceMatrix}.
\end{align}

The bilinear Hamiltonian \eqref{sm:eq:hamilton-matrix} can be diagonalized with $T_{\bm{k}}^{\dagger} H_{\bm{k}, +} T_{\bm{k}} = \mathrm{diag}(E_{\bm{k}, \alpha}, E_{-\bm{k}, \beta})$ and $T_{\bm{k}}^{\dagger} H_{\bm{k}, -} T_{\bm{k}} = \mathrm{diag}(E_{\bm{k}, \beta}, E_{-\bm{k}, \alpha})$, where 
\begin{align}
 	E_{\bm{k}, \alpha} =  \varepsilon_{\bm{k}} + \Delta_{\bm{k}}, \quad E_{\bm{k}, \beta} =  \varepsilon_{\bm{k}} - \Delta_{\bm{k}}, \quad \varepsilon_{\bm{k}} = \sqrt{A_{\bm{k}}^{2} - B_{\bm{k}}^{2} - D_{\bm{k}}^{2}}
  \label{eq:energies}
\end{align}
and \cite{Zyuzin2016}
\begin{align}
 	T_{\bm{k}} = \begin{pNiceMatrix}
	 	\exp(\mathrm{i} \lambda_{\bm{k}}) \cosh \frac{X_{\bm{k}}}{2} & \sinh \frac{X_{\bm{k}}}{2} \\
		\sinh \frac{X_{\bm{k}}}{2} & \exp(-\mathrm{i}  \lambda_{\bm{k}}) \cosh \frac{X_{\bm{k}}}{2}
	\end{pNiceMatrix},
    \label{eq:TmatrixSM}
\end{align}
with 
\begin{align}
    \exp(\mathrm{i}  \lambda_{\bm{k}}) = \frac{-\left(B_{\bm{k}} - \mathrm{i} D_{\bm{k}}\right)}{|B_{\bm{k}} - \mathrm{i} D_{\bm{k}}|} = \frac{-\left(B_{\bm{k}} - \mathrm{i} D_{\bm{k}}\right)}{\sqrt{B_{\bm{k}}^{2} + D_{\bm{k}}^{2}}}, \quad \sinh X_{\bm{k}} = \frac{|B_{\bm{k}} - \mathrm{i} D_{\bm{k}}|}{\varepsilon_{\bm{k}}}, \quad \text{and} \quad \cosh X_{\bm{k}} = \frac{A_{\bm{k}}}{\varepsilon_{\bm{k}}}.
\end{align}

\begin{figure}[]
    \centering       
    \includegraphics[width=9cm]{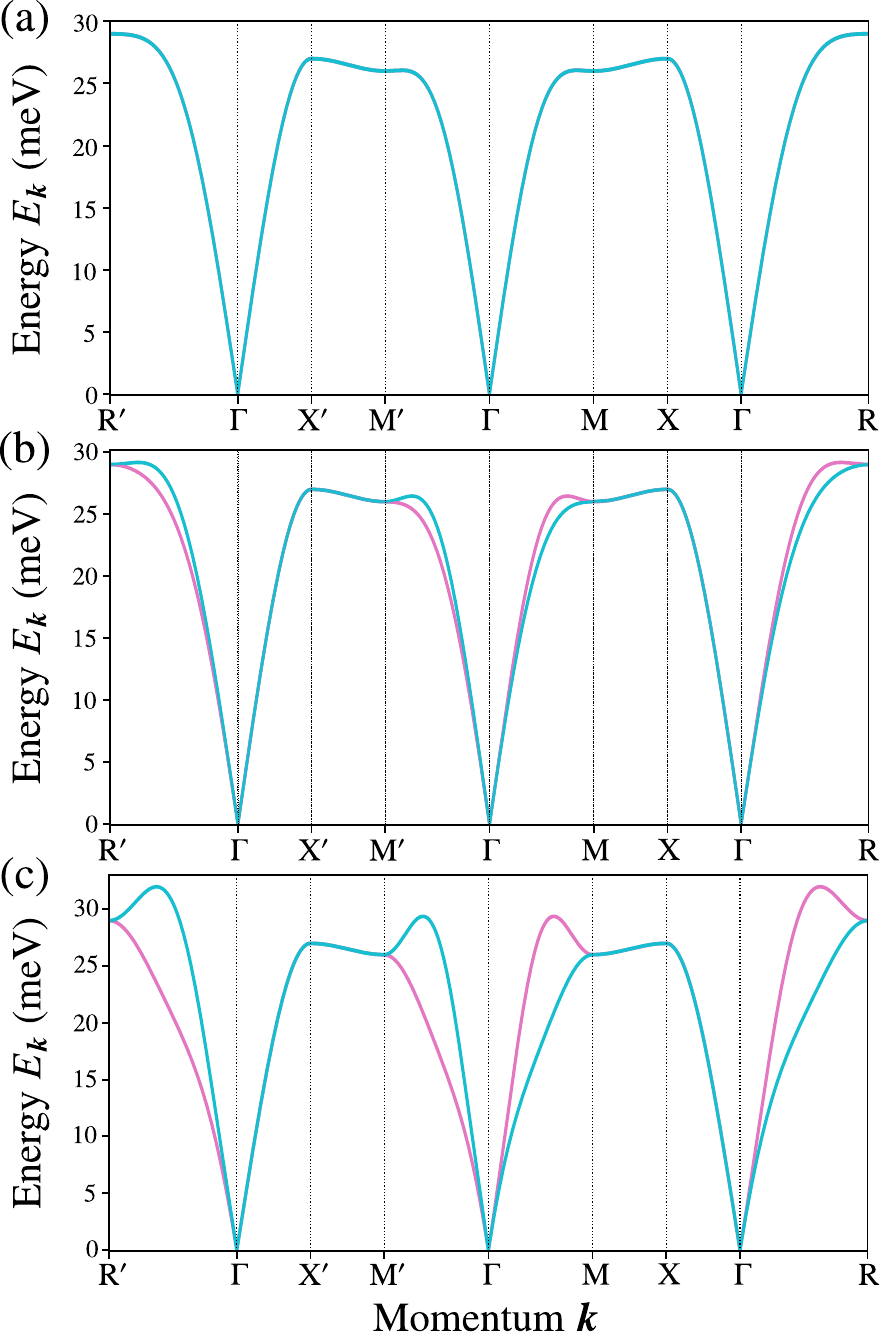}
       \caption{Magnon dispersion for selected values of $\Delta$. The parameters read $S = 5/2$, $J_1 = 1.0$ meV, $J_2 = -0.3$ meV, $J_3 = -0.2$ meV, $D = 0.1$ meV, and $\theta = \phi = \pi/3$. Additionally, we set $\Delta = 0.0$ meV in (a), $\Delta = 0.1$ meV in (b), and $\Delta = 0.5$ meV in (c).}
       \label{fig:bands_delta}
\end{figure}
In Fig.~\ref{fig:bands_delta} we show the dispersion relation of the magnons for selected values of $\Delta$. In Fig.~\ref{fig:bands_delta}(a), we set $\Delta = 0.0$ meV and see a fully degenerate spectrum. In Fig.~\ref{fig:bands_delta}(b), we set $\Delta = 0.1$ meV and we observe the characteristic altermagnetic splitting of the magnon bands. If we increase the value to $\Delta = 0.5$ meV, the splitting becomes more pronounced. 

Since the same $T_{\bm{k}}$ matrix diagonalizes both blocks, the Berry curvature of both bands is identical and is calculated as follows \cite{Zyuzin2016}
\begin{subequations}
\begin{align}
    \Omega^{\gamma}_{\bm{k}} &=  \mathrm{i} \epsilon_{\alpha \beta \gamma} \left( \tau_{3} \frac{\partial T_{\bm{k}}^{\dagger}}{\partial k_{\alpha}} \tau_{3} \frac{\partial T_{\bm{k}}}{\partial k_{\beta}} \right)_{11} \\
	& = \frac{1}{2} \sinh X_{\bm{k}} \left(\partial_{\alpha}\lambda_{\bm{k}} \partial_{\beta} X_{\bm{k}}  - \partial_{\beta} \lambda_{\bm{k}} \partial_{\alpha}X_{\bm{k}}\right).
\end{align}
\end{subequations}
With 
\begin{subequations}
\begin{align}
    \partial_{\alpha}X_{\bm{k}} & =  \frac{\varepsilon_{\bm{k}} \partial_{\alpha} A_{\bm{k}} - A_{\bm{k}}  \partial_{\alpha} \varepsilon_{\bm{k}}}{\varepsilon_{\bm{k}} \sqrt{A_{\bm{k}}^{2} -{\varepsilon_{\bm{k}}^{2}}}}, \\
    \partial_{\alpha} \lambda_{\bm{k}} & =\frac{ D_{\bm{k}} \partial_{\alpha} B_{\bm{k}} - B_{\bm{k}}  \partial_{\alpha} D_{\bm{k}} }{B_{\bm{k}}^{2} + D_{\bm{k}}^{2}},
\end{align}
\end{subequations}
the Berry curvature reads
\begin{align}
    \Omega^{\gamma}_{\bm{k}}  =& \,  \epsilon_{\alpha \beta \gamma} \frac{D_{\bm{k}} \left(\varepsilon_{\bm{k}} \partial_{\beta} A_{\bm{k}} \partial_{\alpha} B_{\bm{k}} + A_{\bm{k}} \partial_{\beta} B_{\bm{k}} \partial_{\alpha} \varepsilon_{\bm{k}} \right) + B_{\bm{k}} \left( \varepsilon_{\bm{k}} \partial_{\beta} D_{\bm{k}} \partial_{\alpha} A_{\bm{k}} + A_{\bm{k}} \partial_{\beta} \varepsilon_{\bm{k}} \partial_{\alpha} D_{\bm{k}} \right)}{2 \varepsilon_{\bm{k}}^{2} \left(B_{\bm{k}}^{2} + D_{\bm{k}}^{2} \right)} \nonumber \\
    = \, & \frac{1}{2 \varepsilon_{\bm{k}}^{2} \left(B_{\bm{k}}^{2} + D_{\bm{k}}^{2} \right)} \left\{D_{\bm{k}} \left[\varepsilon_{\bm{k}} \left(\partial_{\beta} A_{\bm{k}} \partial_{\alpha} B_{\bm{k}} - \partial_{\beta} B_{\bm{k}} \partial_{\alpha} A_{\bm{k}}\right) + A_{\bm{k}} \left(\partial_{\beta} B_{\bm{k}} \partial_{\alpha} \varepsilon_{\bm{k}} - \partial_{\beta} \varepsilon_{\bm{k}} \partial_{\alpha} B_{\bm{k}}\right)\right] \right.\\
	& \left. \qquad \qquad \qquad + B_{\bm{k}}  \left[\varepsilon_{\bm{k}} \left(\partial_{\beta} D_{\bm{k}} \partial_{\alpha} A_{\bm{k}} - \partial_{\beta} A_{\bm{k}} \partial_{\alpha} D_{\bm{k}} \right)+ A_{\bm{k}} \left(\partial_{\beta} \varepsilon_{\bm{k}} \partial_{\alpha} D_{\bm{k}} - \partial_{\beta} D_{\bm{k}} \partial_{\alpha} \varepsilon_{\bm{k}} \right)\right] \right\}.
\end{align}
We find that $\bm{\Omega}_{\bm{k}} = \bm{\Omega}_{-\bm{k}}$ and $\bm{\Omega}_{\bm{k}} \propto D$.

\section{Thermal Hall effect}
\begin{figure}[]
    \centering       \includegraphics[width=7cm]{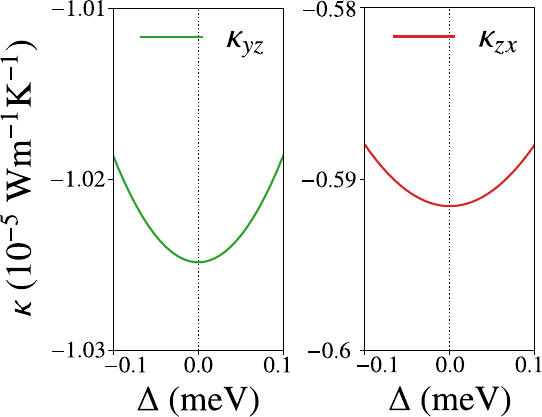}
       \caption{Thermal Hall conductivity as a function of $\Delta$. The parameters read $S = 5/2$, $J_1 = 1.0$ meV, $J_2 = -0.3$ meV, $J_3 = -0.2$ meV, $D = 0.1$ meV, $T = 60$ K, and $\theta = \phi = \pi/3$.}
       \label{fig:THC-delta}
\end{figure}

The thermal Hall conductivity (THC) is computed within the free-boson theory \cite{Matsumoto2011PRL} as
\begin{subequations}
\begin{align}
    \bm{\kappa}_\text{H} & = -\frac{k_{\text{B}}^{2} T }{\hbar V} \sum_{\sigma} \sum_{\bm{k}} \bm{\Omega}_{\bm{k}, \sigma} c_{2}(\rho(E_{\bm{k}, \sigma})) \label{eq:THC-fullthing}\\
    & = -\frac{k_{\text{B}}^{2} T }{\hbar V} \sum_{\bm{k}} \bm{\Omega}_{\bm{k}} \left[c_{2}(\rho(E_{\bm{k}, \alpha})) + c_{2}(\rho(E_{\bm{k}, \beta})) \right] \\
    & = -\frac{k_{\text{B}}^{2} T }{\hbar V} \sum_{\bm{k}} \bm{\Omega}_{\bm{k}} \left[c_{2}(\rho(\varepsilon_{\bm{k}} + \Delta_{\bm{k}})) + c_{2}(\rho(\varepsilon_{\bm{k}} - \Delta_{\bm{k}})) \right] \\
    & \approx -\frac{2 k_{\text{B}}^{2} T }{\hbar V} \sum_{\bm{k}} \bm{\Omega}_{\bm{k}} c_{2}(\rho(\varepsilon_{\bm{k}}))  + O(\Delta^2). \label{eq:THC-expansion}
\end{align}
\end{subequations}
We expanded the THC in small $\Delta$ in Eq.~\eqref{eq:THC-expansion} to show that a finite THC does not require the altermagnetic splitting term $\Delta$. This fact is confirmed in Fig.~\ref{fig:THC-delta}, where we see that even if $\Delta$ reaches zero, there still is a finite THC. 

\section{Symmetries of the thermal conductivity tensor}
The general thermal conductivity tensor reads
\NiceMatrixOptions{cell-space-top-limit = 1pt,cell-space-bottom-limit = 1pt}
\begin{align} \kappa_{ij} =
\begin{pNiceMatrix}[columns-width = 0.55cm]
 \kappa_{xx} & \kappa_{xy} & \kappa_{xz}\\
 \kappa_{yx} & \kappa_{yy} & \kappa_{yz}\\
 \kappa_{zx} & \kappa_{zy} & \kappa_{zz}\\
 \end{pNiceMatrix}. 
\end{align}
Symmetry operations affect the tensor in the following way \cite{Seemann2015}
\begin{align}
 	\kappa_{ij} & = R_{i, i'} R_{j j'} \kappa_{i'j'},
\end{align}
with $R$ being the three-dimensional vector representation of a unitary symmetry of the magnetic point group (MPG), e.g. the unitary mirrors $\mathcal{M}_x, \mathcal{M}_y, \mathcal{M}_z$, where
\begin{align}
    \mathcal{M}_x = \mathrm{diag}(-1, 1, 1), \quad \mathcal{M}_y = \mathrm{diag}(1, -1, 1), \quad \mathcal{M}_z = \mathrm{diag}(1, 1, -1).
    \label{eq:mirrors}
\end{align}
For antiunitary symmetries, like $\mathcal{M}'_x, \mathcal{M}'_y, \mathcal{M}'_z$ (with the prime denoting time reversal), the thermal conductivity tensor has to obey
\begin{align}
    \kappa_{ij} & = R_{i', i} R_{j' j} \kappa_{j'i'}.
\end{align}
In the following Tab.~\ref{tab:symmetries_THC} we show
how the thermal conductivity tensor is restricted by (unitary and antiunitary) mirror symmetries for selected highly symmetric cases with the Néel vector pointing along special directions. For example, in the case $\hat{\bm{N}} \parallel \hat{\bm{z}}$, the magnetic point group contains three mirrors $\mathcal{M}_x$, $\mathcal{M}_y'$, and $\mathcal{M}_z$, which lead to a complete suppression of the thermal Hall effect (no antisymmetric part of the conductivity tensor). In particular, in the main text, we are interested in the thermal Hall conductivity component $\kappa_{zx}$ ($\kappa_{yz}$), which is suppressed when $\hat{\bm{N}} \perp \hat{\bm{x}}$ ($\hat{\bm{N}} \perp \hat{\bm{y}}$).

\begin{table}[H]
\centering
\caption{Restricted shape of the thermal conductivity tensor under the symmetry operations due to (glide) mirrors in the system. Here, $\hat{\bm{N}} \perp \hat{\bm{x}}$ means that $\hat{\bm{N}}$ points in a general direction in the $yz$ plane.}
\label{tab:symmetries_THC}
\begin{tabular}{ p{1.5cm}|p{1.6cm}|p{2.8cm} }
 \toprule
 \centering Orientation of the Néel vector & \centering Existing mirrors in MPG & \centering Effect of mirrors on thermal conductivity tensor $\kappa_{ij}$
  \arraybackslash \\
 \hline

 \centering
 $\hat{\bm{N}} \perp \hat{\bm{x}}$ & \centering $\mathcal{M}_{x}$   & \centering
 $\begin{pNiceMatrix}[columns-width = 0.55cm]
 \kappa_{xx} & 0 & 0\\
 0 & \kappa_{yy} & \kappa_{yz}\\
 0 & \kappa_{zy} & \kappa_{zz}\\
 \end{pNiceMatrix}$  
  \arraybackslash \\
 
 \centering $\hat{\bm{N}} \parallel \hat{\bm{x}}$ & \centering $ \mathcal{M}_{x}', \mathcal{M}_{y}, \mathcal{M}_{z}'$ &   
  \centering
    $\begin{pNiceMatrix}[columns-width = 0.55cm]
 \kappa_{xx}  & 0 & -\kappa_{zx}\\
 0 &  \kappa_{yy} & 0\\
 \kappa_{zx} & 0 & \kappa_{zz}\\
 \end{pNiceMatrix}$
   \arraybackslash  \\
 
\centering  $\hat{\bm{N}} \perp \hat{\bm{y}} $ & \centering $ \mathcal{M}_{y}$ & 
  \centering
 $\begin{pNiceMatrix}[columns-width = 0.55cm]
 \kappa_{xx} & 0 & \kappa_{xz}\\
 0 & \kappa_{yy} & 0\\
 \kappa_{zx} & 0& \kappa_{zz}\\
 \end{pNiceMatrix}$ 
  \arraybackslash \\
 
 \centering $\hat{\bm{N}} \parallel \hat{\bm{y}} $ & \centering $ \mathcal{M}_{x}, \mathcal{M}_{y}', \mathcal{M}_{z}'$   &  
  \centering
$\begin{pNiceMatrix}[columns-width = 0.55cm]
 \kappa_{xx} & 0 & 0\\
 0 & \kappa_{yy} & \kappa_{yz}\\
 0 & -\kappa_{yz} & \kappa_{zz}\\
 \end{pNiceMatrix}$  
   \arraybackslash \\
 
 \centering $\hat{\bm{N}} \perp \hat{\bm{z}} $ & \centering $ \mathcal{M}_{z}'$ & 
  \centering
 $\begin{pNiceMatrix}[columns-width = 0.55cm]
 \kappa_{xx} & 0 & -\kappa_{zx}\\
  0 & \kappa_{yy} & \kappa_{yz}\\
 \kappa_{zx} & -\kappa_{yz} & \kappa_{zz}\\
 \end{pNiceMatrix}$  
  \arraybackslash  \\
 
\centering $\hat{\bm{N}} \parallel \hat{\bm{z}} $ & \centering $  \mathcal{M}_{x}, \mathcal{M}_{y}', \mathcal{M}_{z}$ &
  \centering
$\begin{pNiceMatrix}[columns-width = 0.55cm]
 \kappa_{xx} & 0 & 0\\
 0 & \kappa_{yy} & 0\\
 0 & 0 & \kappa_{zz}\\
\end{pNiceMatrix}$
   \arraybackslash  \\
 \botrule
\end{tabular}
\end{table}

\section{Implementation of strain}
\label{sec:strain}
We apply shear strain in the $[1 1 0]$ direction, which we model by reducing (enhancing) the nearest-neighbor exchange coupling by $\delta J_1$ for bonds with a finite (zero) projection onto the $[1 1 0]$ direction. 
This procedure changes the element $B_{\bm{k}}$ of the Hamilton matrix as follows
\begin{align}
	B_{\bm{k}} \rightarrow B_{\bm{k}}' =\,&  2 S \left[ \left( J_{1} + \delta J_1 \right) \left( \cos \frac{k_{x} + k_{y} + k_{z}}{2} + \cos \frac{k_{x} + k_{y} - k_{z}}{2} \right) + \left( J_{1} - \delta J_1 \right) \left( \cos \frac{k_{x} - k_{y} + k_{z}}{2} + \cos \frac{-k_{x} + k_{y} + k_{z}}{2} \right)\right].
\label{eq:strain}
\end{align}
This leads to a different Berry curvature from which we can compute the THC depending on the strain parameter $\delta J_1$, as done in the main text.

%
%
\section{Magnetization induced by thermal fluctuations}
We have found that the classical order is fully compensated, but supports a thermal Hall effect. Consequently, there must also be a finite magnetization. Indeed, we show here that thermal fluctuations can lead to a finite but tiny magnetization within linear spin-wave theory. We calculate the thermodynamic expectation value of the magnetization in the free theory as follows,
\begin{subequations}
\begin{align}
    \frac{\langle \bm{M} \rangle_{2}}{ S} & = \frac{1}{S}\langle \bm{S}^{\text{A}} + \bm{S}^{\text{B}} \rangle_{2} \label{eq:magnetization_ansatz} \\
	& = \Big\langle \frac{1}{N S} \sum_{i} \left( \bm{S}_{i}^{\text{A}} + \bm{S}_{i}^{\text{B}} \right) \Big\rangle_{2}  \\
	& = \frac{\hat{\bm{N}}}{N S} \sum_{i} \left( \langle b_{i}^{\dagger}b_{i} \rangle_{2} - \langle a_{i}^{\dagger}a_{i} \rangle_{2} \right)  \\
	& = \frac{\hat{\bm{N}}}{N S} \sum_{\bm{k}} \left( \langle b_{\bm{k}}^{\dagger}b_{\bm{k}} \rangle_{2} - \langle a_{\bm{k}}^{\dagger}a_{\bm{k}} \rangle_{2} \right).
\end{align}
\end{subequations}
We know that
\begin{align}
    H_2 = \frac{1}{2} \sum_{\bm{k}} \sum_{ \sigma = \pm} \underbrace{\bm{\Psi}_{\bm{k}, \sigma}^{\dagger} \left(T_{\bm{k}}^{\dagger}\right)^{-1}}_{\bm{\Phi}_{\bm{k}, \sigma}^{\dagger}} 
    T_{\bm{k}}^{\dagger} H_{\bm{k}, \sigma} T_{\bm{k}}
    \underbrace{\left(T_{\bm{k}}\right)^{-1} \bm{\Psi}_{\bm{k}}}_{\bm{\Phi}_{\bm{k}, \sigma}},
\end{align}
with $T_{\bm{k}}$ given in Eq.~\eqref{eq:TmatrixSM}.
Using $\bm{\Psi}_{\bm{k}, \sigma}^{\dagger} = \bm{\Phi}_{\bm{k}, \sigma}^{\dagger}T_{\bm{k}}^{\dagger} $, we find 
\begin{subequations}
\begin{align}
    \langle a_{\bm{k}}^{\dagger}a_{\bm{k}} \rangle_{2} &= \frac{1}{2} \left[ \left( \cosh X_{\bm{k}} - 1 \right) \langle \beta_{-\bm{k}}\beta_{-\bm{k}}^{\dagger}\rangle_{2} + \left( \cosh X_{\bm{k}} + 1 \right)\langle \alpha_{\bm{k}}^{\dagger}\alpha_{\bm{k}}\rangle_{2} + \sinh X_{\bm{k}} \mathrm{e}^{- \mathrm{i} \lambda{\bm{k}}} \left( \mathrm{e}^{2 \mathrm{i} \lambda{\bm{k}}} \langle\beta_{-\bm{k}} \alpha_{\bm{k}} \rangle_{2} +\langle \alpha_{\bm{k}}^{\dagger} \beta_{-\bm{k}}^{\dagger}\rangle_{2} \right) \right], \\
	\langle b_{\bm{k}}^{\dagger}b_{\bm{k}} \rangle_{2} &= \frac{1}{2} \left[ \left( \cosh X_{\bm{k}} - 1 \right) \langle \alpha_{-\bm{k}}\alpha_{-\bm{k}}^{\dagger}\rangle_{2} + \left( \cosh X_{\bm{k}} + 1 \right)\langle \beta_{\bm{k}}^{\dagger}\beta_{\bm{k}}\rangle_{2} + \sinh X_{\bm{k}} \mathrm{e}^{- \mathrm{i} \lambda{\bm{k}}} \left( \mathrm{e}^{2 \mathrm{i} \lambda{\bm{k}}} \langle\beta_{\bm{k}}^{\dagger} \alpha_{-\bm{k}}^{\dagger} \rangle_{2} + \langle \alpha_{-\bm{k}} \beta_{\bm{k}}\rangle_{2}\right) \right].
\end{align}
\end{subequations}
With the commutation relations $ [ \alpha_{\bm{k}}, \alpha_{\bm{k}'}^{\dagger} ] = \delta_{{\bm{k}}, {\bm{k}}'}$, $ [ \alpha_{\bm{k}}^{\dagger}, \alpha_{{\bm{k}}'}^{\dagger} ] = [ \alpha_{\bm{k}}, \alpha_{\bm{k}'} ]  = 0$, as well as $\langle \alpha_{\bm{k}}^{\dagger}\alpha_{\bm{k}} \rangle_{2} \approx \rho(E_{\bm{k},\alpha})$, which is the Bose-Einstein function, and $\langle\beta_{-\bm{k}} \alpha_{\bm{k}} \rangle_{2} = \langle \alpha_{\bm{k}}^{\dagger} \beta_{-\bm{k}}^{\dagger}\rangle_{2} = \langle\beta_{\bm{k}}^{\dagger} \alpha_{-\bm{k}}^{\dagger} \rangle_{2} = \langle \alpha_{-\bm{k}} \beta_{\bm{k}}\rangle_{2} = 0$, we get
\begin{align}
    \frac{ \langle \bm{M} \rangle_2}{S} & = \frac{\hat{\bm{N}}}{N S} \sum_{\bm{k}} \left[\rho(E_{\bm{k}, \beta}) - \rho(E_{\bm{k}, \alpha})\right].
\label{eq:magnetisation}
\end{align}
Thus, while quantum fluctuations drop out, thermal fluctuations can give rise to a finite magnetization.

\begin{figure}[]
    \centering       \includegraphics[width=15cm]{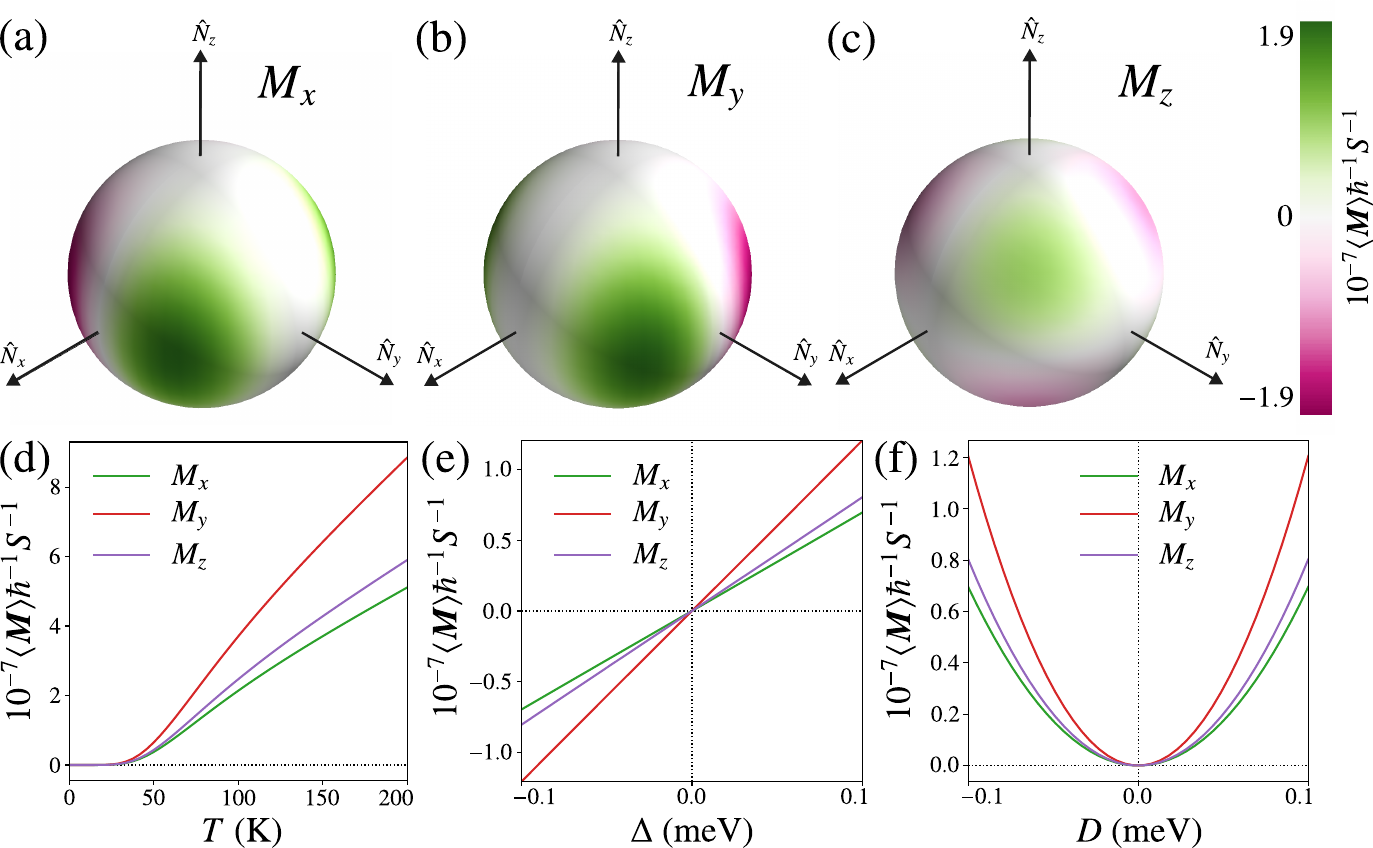}
       \caption{(a)-(c) Néel vector dependence of the magnetization components. (d) Temperature dependence of the magnetization. (e) Linear dependence of the magnetization on the altermagnetic splitting term $\Delta$. (f) Quadratic dependence of the magnetization on the Dzyaloshinskii-Moriya interaction. We used $\theta = \phi = \pi/3$ for (d)-(f). Other parameters read $S = 5/2$, $J_1 = 1.0$ meV, $J_2 = -0.3$ meV, $J_3 = -0.2$ meV, $\Delta = D = 0.1$ meV, and  $T = 60$ K. Note that we have recovered $\hbar$.}
       \label{fig:mag}
\end{figure}

In Fig.~\ref{fig:mag}, we have evaluated Eq.~\eqref{eq:magnetisation} numerically and find that thermal fluctuations indeed induce a finite, albeit tiny, magnetization. Figures~\ref{fig:mag}(a-c) show the Néel vector dependence of the $x$, $y$ and $z$ components of the magnetization. We note that the magnetization vanishes for $\hat{\bm{N}} \parallel \hat{\bm{x}}, \hat{\bm{y}}, \hat{\bm{z}}$ for all three components, which is a combined result of magnetic point group symmetries and the fact that our calculation only captures the magnetization \emph{along} the Néel vector. In addition, Fig.~\ref{fig:mag}(d) shows how the finite magnetization becomes thermally activated. 

Next, we explore how the magnetization depends on the altermagnetic splitting term $\Delta$ and the DMI. We approximate
\begin{subequations}
\begin{align}
    \frac{\langle \bm{M} \rangle_{2}}{  S} & =  \frac{\hat{\bm{N}}}{N S} \sum_{\bm{k}} \left[\rho (\varepsilon_{\bm{k}} - \Delta_{\bm{k}}) - \rho_(\varepsilon_{\bm{k}} + \Delta_{\bm{k}})\right]\\
    &= \frac{\hat{\bm{N}}}{N S} \sum_{\bm{k}} \left( 2 \Delta_{\bm{k}} \right) \frac{\rho(\varepsilon_{\bm{k}} - \Delta_{\bm{k}}) - \rho(\varepsilon_{\bm{k}} + \Delta_{\bm{k}})}{2 \Delta_{\bm{k}}}\\
    & \approx - \frac{2\hat{\bm{N}}}{N S} \sum_{\bm{k}}  \Delta_{\bm{k}}  \frac{\partial \rho(\varepsilon)}{\partial \varepsilon}\Big|_{\varepsilon =  \varepsilon_{\bm{k}}}, \label{eq:mag-approx}
\end{align}
\end{subequations}
from where we can see the linear dependence on the altermagnetic splitting $\Delta$, which matches the behaviour seen in Fig.~\ref{fig:mag}(e).
Furthermore, we know that 
\begin{align}
    \frac{\partial \rho(\varepsilon)}{\partial \varepsilon} = -\frac{\exp \left(\varepsilon_{\bm{k}} \beta \right) \beta}{\left[\exp \left(\varepsilon_{\bm{k}} \beta \right) -1 \right]^2}, \quad \text{with } \beta = \left(k_\text{B} T\right)^{-1}.
\end{align}
So we expand
\begin{align}
    F = \frac{\partial \rho(\varepsilon)}{\partial \varepsilon}\Big|_{\varepsilon = \varepsilon_{\bm{k}}}  = F^{(0)} + F^{(1)} + F^{(2)} + O(D^3),
\end{align}
where $F^{(n)} = O(D^n)$, explicitly
\begin{subequations}
\begin{align}
    F^{(0)} &=  - \frac{\exp\left(\sqrt{A_{\bm{k}}^2 - B_{\bm{k}}^2} \beta \right) \beta}{\left[ \exp\left(\sqrt{A_{\bm{k}}^2 - B_{\bm{k}}^2} \beta \right) -1 \right]^2} \propto D^0, \\
     F^{(1)} &=  0, \\
     F^{(2)} & = - \frac{ \exp\left(\sqrt{A_{\bm{k}}^2 - B_{\bm{k}}^2} \beta \right) \left[1+\exp\left(\sqrt{A_{\bm{k}}^2 - B_{\bm{k}}^2} \beta \right)\right] \beta^2}{2 \sqrt{A_{\bm{k}}^2 - B_{\bm{k}}^2} \left[\exp\left(\sqrt{A_{\bm{k}}^2 - B_{\bm{k}}^2} \beta \right) -1 \right]^3} D^2_{\bm{k}}  \propto D^2.
\end{align}
\end{subequations}
The magnetization coming from the $F^{(0)}$ term vanishes in Eq.~\eqref{eq:mag-approx} because of its $C_{4z}$ symmetry, making $D^2$ the leading order, which is in accordance with Fig.~\ref{fig:mag}(f).

\begin{figure}[]
\centering
       \includegraphics[width=15cm]{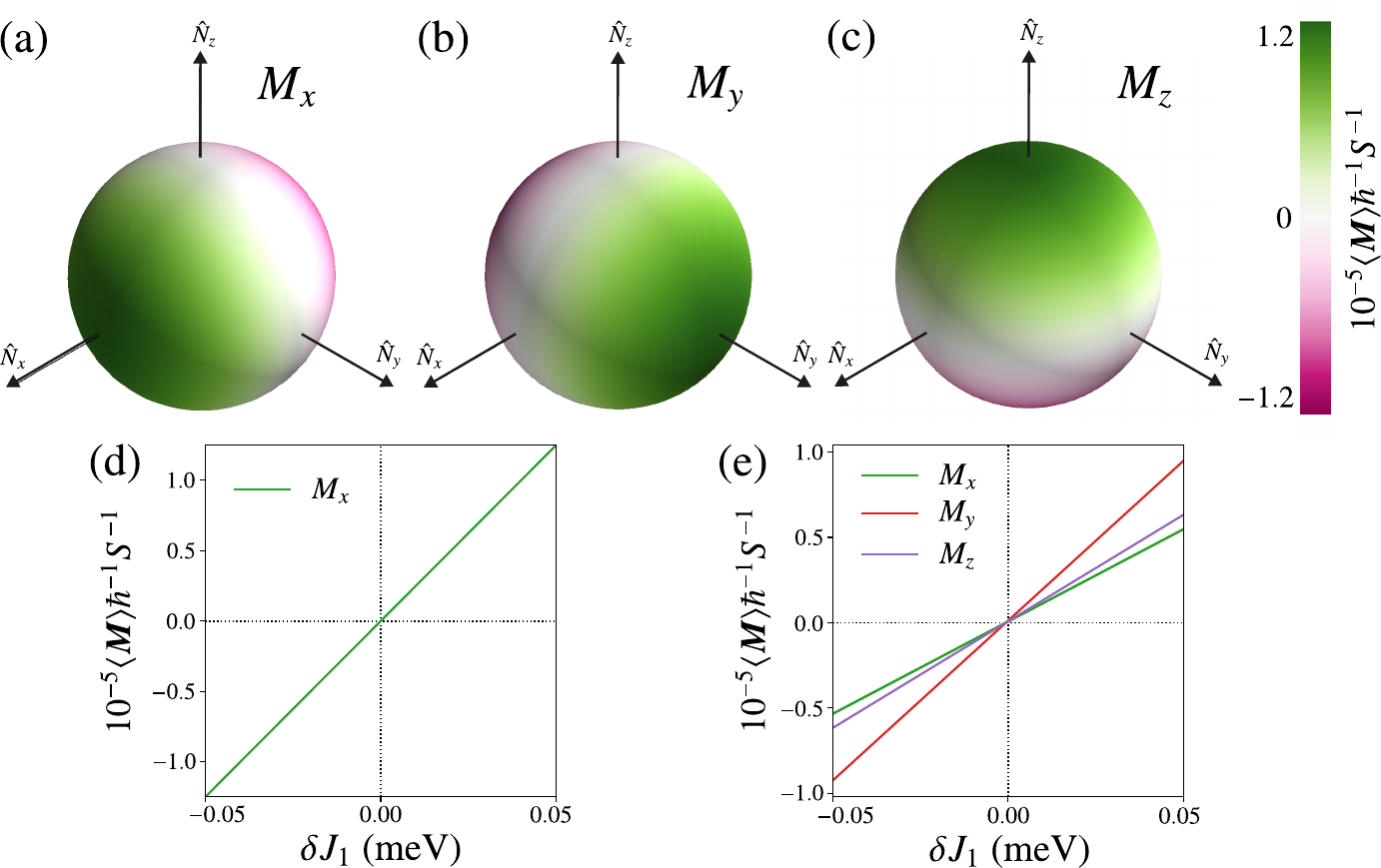}
       \caption{Effects of strain on the magnetization without spin-orbit coupling ($D = 0.0$ meV). (a)-(c) Néel vector dependence of the magnetization components. (d) Strain dependence of the component $M_x$ for $\hat{\bm{N}} \parallel \hat{x}$. (e) Strain dependence for $\theta = \phi = \pi/3$. All other parameters as in Fig.~\ref{fig:mag}. Note that we have recovered $\hbar$.}
       \label{fig:mag_strain}
\end{figure}

Finally, we also investigate the thermal fluctuation induced piezomagnetism without spin-orbit coupling ($D = 0.0$ meV). 
In Fig.~\ref{fig:mag_strain}(a-c) the application of strain [implemented as in Eq.~\eqref{eq:strain}, replacing $B_{\bm{k}}$ by $B'_{\bm{k}}$] shows an increase of the magnetization by two orders of magnitude in comparison with Fig.~\ref{fig:mag}(a-c). This is because the $C_{4z}$ symmetry of $F^{(0)}$ is broken when replacing $B_{\bm{k}}$ by $B'_{\bm{k}}$, rendering $F^{(0)}$ the leading contribution to the magnetization. This finite magnetization is linear in strain, as shown in Fig.~\ref{fig:mag_strain}(d) for $M_x$ in the special case of $\hat{\bm{N}} \parallel \hat{\bm{x}}$, and in Fig.~\ref{fig:mag_strain}(e) for all magnetization components for a general direction of the Néel vector.

%
%

\section{Spin Nernst conductivity}
The spin Nernst conductivity (SNC) reads \cite{Cheng2016, Zyuzin2016}
\begin{subequations}
\begin{align}
    \bm{\alpha}^\gamma = 
    \begin{pmatrix}
        \alpha^\gamma_{yz} \\  \alpha^\gamma_{zx} \\  \alpha^\gamma_{xy}
    \end{pmatrix}
    & = -\frac{k_\text{B}}{\hbar V} \sum_{n, \bm{k}}  \langle s^\gamma \rangle_{n, \bm{k}} \Omega_{n, \bm{k}} c_1(\rho(E_{n, \bm{k}})) \\
    & = -\frac{k_\text{B} \hat{N}^\gamma}{ V} \sum_{\bm{k}} \Omega_{\bm{k}} \left[c_1(\rho(E_{\bm{k},\beta})) - c_1(\rho(E_{\bm{k}, \alpha}))\right] \\
    & \approx -\frac{k_\text{B}  \hat{N}^\gamma}{V} \sum_{\bm{k}} \Omega_{\bm{k}}  \left(2 \Delta_{\bm{k}} \right) \frac{c_1(\rho(\varepsilon_{\bm{k}} - \Delta_{\bm{k}})) - c_1(\rho(\varepsilon_{\bm{k}} + \Delta_{\bm{k}}))}{2 \Delta_{\bm{k}}} \\
     & \approx \frac{2 k_\text{B}  \hat{N}^\gamma}{V} \sum_{\bm{k}} \Omega_{\bm{k}}  \Delta_{\bm{k}} \frac{\partial c_1(\rho(x))}{\partial x}\Big|_{x = \varepsilon_{\bm{k}}} + O(\Delta^3),
\end{align}
\end{subequations}
where $\gamma = x,y,z$ describes the transported spin component. We used 
$\langle s^\gamma \rangle_{\bm{k}, \alpha} = -\hat{N}^\gamma \hbar$ and $\langle s^\gamma \rangle_{\bm{k}, \beta} = +\hat{N}^\gamma \hbar$, 
and expanded in small $\Delta$. This expansion shows that the SNC depends both on the DMI $D$ and the altermagnetic splitting term $\Delta$.

The general spin Nernst conductivity tensor reads
\begin{align}
 \alpha_{ij}^{\gamma} =
\begin{pNiceMatrix}
 \alpha_{xx}^{\gamma} & \alpha_{xy}^{\gamma} & \alpha_{xz}^{\gamma}\\
 \alpha_{yx}^{\gamma} & \alpha_{yy}^{\gamma} & \alpha_{yz}^{\gamma}\\
 \alpha_{zx}^{\gamma} & \alpha_{zy}^{\gamma} & \alpha_{zz}^{\gamma}\\
 \end{pNiceMatrix},
\end{align}
and symmetry operations affect it in the following way \cite{Seemann2015}
\begin{align}
     \alpha_{ij}^{\gamma} & = \mathrm{det}(R)R_{\gamma, \gamma'} R_{i i'} R_{j j'} \alpha_{i' j'}^{\gamma'},
\end{align}
where $R$ is the three-dimensional vector representation of a symmetry that can be both unitary or antiunitary, e.g. as in Eq.~\eqref{eq:mirrors}.
In Tab.~\ref{tab:symmetries_SNC} we show the spin Nernst conductivity tensor under the effect of the mirrors contained in the respective magnetic point group for selected highly-symmetric directions of the Néel vector.

In the main text, we present data for $\alpha^x_{yz}$ in Fig.~3(c), an element that is never suppressed by symmetries. Instead, the zeros of $\alpha^x_{yz}$ for $\hat{\bm{N}} \perp \hat{\bm{x}}$ are the result of the magnon spin expectation value being parallel to $\hat{\bm{N}}$, such that no $x$ component of the spin can be transported. In Fig.~3(d) of the main text, we plot $\alpha^x_{zx}$, which according to Tab.~\ref{tab:symmetries_SNC}, can only be nonzero if $\hat{\bm{N}}$ is neither perpendicular to $\hat{\bm{x}}$ nor $\hat{\bm{y}}$.

\begin{table}[H]
\centering
\caption{Spin Nernst conductivity tensor under the symmetry operations due to (glide) mirrors in the system. Here, $\hat{\bm{N}} \perp \hat{\bm{x}}$ means that $\hat{\bm{N}}$ points in a general direction in the $yz$ plane.}
\label{tab:symmetries_SNC}
\begin{tabular}{ p{1.5cm}  | p{1.6cm} | p{6.cm}}
 \toprule
  \centering Orientation of the Néel vector  &  \centering Existing mirrors in MPG &  \centering Effect of mirrors on spin Nernst conductivity tensor $\alpha_{ij} $
 \arraybackslash \\

\hline
 \centering $\hat{\bm{N}} \perp \hat{\bm{x}} $ & \centering $ \mathcal{M}_{x}$   & 
\centering 
 $\begin{pNiceMatrix}[columns-width = 0.4cm]
 \alpha_{xx}^{x} & 0 & 0\\
 0 & \alpha_{yy}^{x} & \alpha_{yz}^{x}\\
 0 & \alpha_{zy}^{x} & \alpha_{zz}^{x}\\
 \end{pNiceMatrix}$,
 $\begin{pNiceMatrix}[columns-width = 0.4cm]
0 & \alpha_{xy}^{y} & \alpha_{xz}^{y}\\
 \alpha_{yx}^{y} & 0 & 0\\
 \alpha_{zx}^{y} & 0 & 0\\
 \end{pNiceMatrix}$,
 $\begin{pNiceMatrix}[columns-width = 0.4cm]
 0 & \alpha_{xy}^{z} & \alpha_{xz}^{z}\\
 \alpha_{yx}^{z} & 0& 0\\
 \alpha_{zx}^{z} & 0 & 0\\
 \end{pNiceMatrix}$
 \arraybackslash \\
\centering
 $\hat{\bm{N}} \parallel \hat{\bm{x}} $ & \centering $ \mathcal{M}_{x}', \mathcal{M}_{y}, \mathcal{M}_{z}'$ & 
 \centering 
$ \begin{pNiceMatrix}[columns-width = 0.4cm]
 0  & 0  & 0\\
 0  & 0  & \alpha_{yz}^{x}\\
 0 & \alpha_{zy}^{x} & 0 \\
 \end{pNiceMatrix},
 \begin{pNiceMatrix}[columns-width = 0.4cm]
0  & 0  & \alpha_{xz}^{y}\\
 0  & 0  & 0\\
 \alpha_{zx}^{y} & 0 & 0 \\
 \end{pNiceMatrix},
 \begin{pNiceMatrix}[columns-width = 0.4cm]
 0 & \alpha_{xy}^{z} & 0\\
 \alpha_{yx}^{z} & 0& 0\\
 0 & 0 & 0\\
 \end{pNiceMatrix}
 $ 
\arraybackslash \\
\centering
 $\hat{\bm{N}} \perp \hat{\bm{y}} $ &\centering  $ \mathcal{M}_{y}$ & 
 \centering
  $\begin{pNiceMatrix}[columns-width = 0.4cm]
 0 & \alpha_{xy}^{x} & 0\\
 \alpha_{yx}^{x} & 0 & \alpha_{yz}^{x}\\
0 & \alpha_{zy}^{x} & 0\
 \end{pNiceMatrix}$,
 $\begin{pNiceMatrix}[columns-width = 0.4cm]
 \alpha_{xx}^{y} & 0 & \alpha_{xz}^{y}\\
 0 & \alpha_{yy}^{y} & 0\\
 \alpha_{zx}^{y} & 0 & \alpha_{zz}^{y}\\
 \end{pNiceMatrix}$,
 $\begin{pNiceMatrix}[columns-width = 0.4cm]
 0 & \alpha_{xy}^{z} & 0\\
 \alpha_{yx}^{z} & 0 & \alpha_{yz}^{z}\\
 0 & \alpha_{zy}^{z} & 0\\
 \end{pNiceMatrix}$
 \arraybackslash \\

\centering
 $\hat{\bm{N}} \parallel \hat{\bm{y}} $ & \centering $ \mathcal{M}_{x}, \mathcal{M}_{y}', \mathcal{M}_{z}'$   & 
 \centering
$\begin{pNiceMatrix}[columns-width = 0.4cm]
 0  & 0  & 0\\
 0  & 0  & \alpha_{yz}^{x}\\
 0 & \alpha_{zy}^{x} & 0 \\
 \end{pNiceMatrix},
 \begin{pNiceMatrix}[columns-width = 0.4cm]
0  & 0  & \alpha_{xz}^{y}\\
 0  & 0  & 0\\
 \alpha_{zx}^{y} & 0 & 0 \\
 \end{pNiceMatrix},
 \begin{pNiceMatrix}[columns-width = 0.4cm]
 0 & \alpha_{xy}^{z} & 0\\
 \alpha_{yx}^{z} & 0& 0\\
 0 & 0 & 0\\
 \end{pNiceMatrix}
 $  \arraybackslash \\

\centering
 $\hat{\bm{N}} \perp \hat{\bm{z}} $ & \centering $ \mathcal{M}_{z}'$ 
  & 
  \centering
  $\begin{pNiceMatrix}[columns-width = 0.4cm]
 0  & 0  & \alpha_{xz}^{x}\\
 0  & 0  & \alpha_{yz}^{x}\\
 \alpha_{zx}^{x} & \alpha_{zy}^{x} & 0 \\
 \end{pNiceMatrix}$,
 $\begin{pNiceMatrix}[columns-width = 0.4cm]
0  & 0  & \alpha_{xz}^{y}\\
 0  & 0  & \alpha_{yz}^{y}\\
 \alpha_{zx}^{y} & \alpha_{zy}^{y} & 0 \\
 \end{pNiceMatrix}$,
 $\begin{pNiceMatrix}[columns-width = 0.4cm]
 \alpha_{xx}^{z} & \alpha_{xy}^{z} & 0 \\
 \alpha_{yx}^{z} & \alpha_{yy}^{z} & 0 \\
 0  & 0  & \alpha_{zz}^{z}\\
 \end{pNiceMatrix}$  
  \arraybackslash \\
  
\centering
 $\hat{\bm{N}} \parallel \hat{\bm{z}}$ & \centering $\mathcal{M}_{x}, \mathcal{M}_{y}', \mathcal{M}_{z}$ &
 \centering
   $ \begin{pNiceMatrix}[columns-width = 0.4cm]
 0  & 0  & 0\\
 0  & 0  & \alpha_{yz}^{x}\\
 0 & \alpha_{zy}^{x} & 0 \\
 \end{pNiceMatrix},
 \begin{pNiceMatrix}[columns-width = 0.4cm]
0  & 0  & \alpha_{xz}^{y}\\
 0  & 0  & 0\\
 \alpha_{zx}^{y} & 0 & 0 \\
 \end{pNiceMatrix},
 \begin{pNiceMatrix}[columns-width = 0.4cm]
 0 & \alpha_{xy}^{z} & 0\\
 \alpha_{yx}^{z} & 0& 0\\
 0 & 0 & 0\\
 \end{pNiceMatrix}
 $  \arraybackslash \\
 \botrule
\end{tabular}

\end{table}

%
%

\section{Spurious symmetries of the harmonic theory and magnon-magnon interactions}
In spin-wave theory, it is established that both the ground state energy $H_0$ \cite{Rau2018PseudoGoldstone} and the harmonic theory $H_2$ \cite{Gohlke2023PRL} can exhibit spurious symmetries, which lead to pseudo-Goldstone modes and band degeneracies, respectively. The present model is an example of both, but we emphasize that the spin Hamiltonian does not include all magnetic interactions allowed by the lattice symmetries. Thus, the problems discussed here are a feature of the simplicity of the model and we address them only for completeness. It is expected that an inclusion of local magnetocrystalline anisotropies caused by the octahedral environment of the magnetic ions immediately rectifies the linear spin-wave theory.

First, note that the DMI does not enter $H_0$ because $\sum_j \bm{D}_{\delta_j} = 0$ and there is a spurious continuous spin-rotation symmetry for the classical ground state, leading to two pseudo-Goldstone modes. Standard order-by-disorder arguments lift this degeneracy and select a preferred orientation of the N\'{e}el vector, and magnon-magnon interactions will gap out the pseudo-Goldstone modes \cite{Rau2018PseudoGoldstone}.

Second, our main focus is on the spurious symmetries of $H_2$. Since the harmonic Hamiltonian only contains the components of the DMI vectors parallel to $\bm{N}$, that is,
\begin{align}
    \bm{D}_\delta \cdot \hat{\bm{N}} 
    =
    \begin{pmatrix}
            D_{\delta,x} \\ D_{\delta,y} \\ 0
        \end{pmatrix}
    \cdot 
    \begin{pmatrix}
            N_x \\ N_y \\ N_z
    \end{pmatrix}
    =
    N_x D_{\delta,x} + N_y D_{\delta,y},
    \label{eq:dmiinh2}
\end{align}
it is blind to the spin-rotation symmetry breaking of the perpendicular components of the DMI vector. As a result, together with the collinear ground state order, the Hamilton kernel of $H_2$ is block diagonal, spin is a good quantum number, and the $\alpha$ and $\beta$ magnons are degenerate for $\Delta = 0$, as already mentioned in the main text. In addition, the harmonic theory $H_2$ shows a spurious $C'_{2 z} = T C_{2 z}$ symmetry, where $T$ denotes time-reversal symmetry and $C_{2 z}$ a two-fold rotation symmetry about the $z$ direction. To appreciate that this symmetry is spurious, let's apply it to the Néel vector $\bm{N}$,
\begin{align}
    C'_{2z} \bm{N} 
    =
    T C_{2z} \bm{N} 
    = T C_{2z} 
        \begin{pmatrix}
            N_x \\ N_y \\ N_z
        \end{pmatrix}
    = T 
        \begin{pmatrix}
            -N_x \\ -N_y \\ N_z
        \end{pmatrix}
    = 
        \begin{pmatrix}
            N_x \\ N_y \\ -N_z
        \end{pmatrix}
    \ne \bm{N}.
        \label{eq:C2zNeel}
\end{align}
Clearly, it is not a good symmetry of the system, because the $z$ component of $\bm{N}$ has flipped. However, according to Eq.~\eqref{eq:dmiinh2}, the very component of $\bm{N}$ that according to Eq.~\eqref{eq:C2zNeel} renders $T C_{2z}$ not a good symmetry of the system is not captured by the harmonic theory. This is because the DMI vectors have zero $z$ components. As a result, $T C_{2z}$ is an accidental or spurious symmetry of the harmonic theory.

Let us study the implications of $C'_{2z}$ for the thermal Hall conductivity. We study how both the Berry curvature and the magnon dispersion transform under $C'_{2z}$:
\begin{align}
    T C_{2z} \bm{\Omega}_{\bm{k}} 
    &= T C_{2z} 
        \begin{pmatrix}
            \Omega_x \\ \Omega_y \\ \Omega_z
        \end{pmatrix}(k_x,k_y,k_z)
    = T  
        \begin{pmatrix}
            -\Omega_x \\ -\Omega_y \\ \Omega_z
        \end{pmatrix}(-k_x,-k_y,k_z)
    =  
        \begin{pmatrix}
            \Omega_x \\ \Omega_y \\ -\Omega_z
        \end{pmatrix}(k_x,k_y,-k_z),
    \\
    T C_{2z} E_{\bm{k}} 
    &= 
    T C_{2z} E(k_x,k_y,k_z)
    =
    T E(-k_x,-k_y,k_z)
    =
    E(k_x,k_y,-k_z).
\end{align}
In particular, it follows that 
\begin{align}
    \Omega_z(k_x,k_y,k_z) c_2[\rho(E(k_x,k_y,k_z))] = - \Omega_z(k_x,k_y,- k_z) c_2[\rho(E(k_x,k_y,-k_z))],
\end{align}
such that the integral over all momenta in Eq.~\eqref{eq:THC-fullthing} leads to a cancellation and a net zero $\kappa_{xy}$. We conclude that the spurious symmetry enforces $\kappa_{xy} = 0$ in the harmonic theory. A similar argument applies to $\alpha^\gamma_{xy} = 0$.

Next, we will argue why magnon-magnon interactions lift all spurious symmetries.


%
%
\subsection{Derivation of three-magnon interactions}
\label{sec:mminteractions}
To include the three-magnon interactions, we must derive the contribution of the DMI to $H_3$. As we have found in Eq.~\eqref{eq:H1-H3}, the Hamiltonian piece $H_{D}^{\perp \hat{\bm{N}}}$ leads to $H_1$ and $H_3$ terms. We obtain 
\begin{align}
        H_3^D 
    = &\,\frac{1}{2} \sqrt{\frac{S}{2}}\sum_{i \in \text{A}} \sum_{\delta} \left( D_\delta^+ b_{i + \delta}^\dagger a_i  b_{i + \delta} + D_\delta^- a_i^\dagger b_{i + \delta}^\dagger b_{i + \delta} +  D_\delta^+ b_{i + \delta}^\dagger a_i^\dagger  a_i + D_\delta^- a_i^\dagger b_{i + \delta} a_i \right) \nonumber \\
     & + \frac{1}{2} \sqrt{\frac{S}{2}}\sum_{i \in \text{B}} \sum_{\delta} \left( D_\delta^+ b_i^\dagger a_{i + \delta}^\dagger a_{i + \delta}  + D_\delta^- a_{i + \delta}^\dagger b_i  a_{i + \delta} +  D_\delta^+ b_i^\dagger a_{i + \delta}  b_i + D_\delta^- a_{i + \delta}^\dagger b_i^\dagger b_i \right),
\end{align}
where we have defined
\begin{align}
    D^{\pm}_{\delta} = - \bm{D}_{\delta} \cdot  \left(\hat{\bm{y}} \pm \mathrm{i} \hat{\bm{x}} \right).
\end{align}
After a Fourier transformation~\eqref{eq:FT} the three-magnon contribution reads
\begin{align}
    H_3^D
    & = \sum_{ \bm{k}, \bm{q}} \left( D_{\bm{k}}^- a_{\bm{k}}^\dagger b_{\bm{q}}^\dagger  b_{\bm{k}+\bm{q}}  + D_{\bm{k}}^+ b_{\bm{k}}^\dagger a_{\bm{q}}^\dagger  a_{\bm{k}+\bm{q}} + \mathrm{H.c.}\right),
\end{align}
with
\begin{align}
    D_{\bm{k}}^\pm = \sqrt{\frac{S}{2 N}} \sum_{\delta} D_\delta^\pm \cos (\bm{k} \cdot \bm{\delta}).
\end{align}
We perform a Bogoliubov transformation as follows
\begin{align}
    \begin{pNiceMatrix}
        a_{\bm{k}}^\dagger  \\ b_{-\bm{k}} \\ b_{\bm{k}}^\dagger  \\ a_{-\bm{k}}
    \end{pNiceMatrix} 
    =
    \begin{pNiceMatrix}
        u_{\bm{k}}^* & v_{\bm{k}} & 0 & 0 \\
        v_{\bm{k}} & u_{\bm{k}} & 0 & 0 \\
        0 & 0 & u_{\bm{k}}^* & v_{\bm{k}} \\
        0 & 0 & v_{\bm{k}} & u_{\bm{k}}
    \end{pNiceMatrix}
    \begin{pNiceMatrix}
        \alpha_{\bm{k}}^\dagger \\ \beta_{-\bm{k}} \\ \beta_{\bm{k}}^\dagger \\ \alpha_{-\bm{k}}
    \end{pNiceMatrix},
\end{align}
where
\begin{align}
    u_{\bm{k}} = \mathrm{e}^{\mathrm{i} \lambda_{\bm{k}}} \cosh \frac{X_{\bm{k}}}{2}, \qquad v_{\bm{k}} = \sinh \frac{X_{\bm{k}}}{2},
\end{align}
and thus find 
\begin{subequations}
\begin{align}
    a_{\bm{k}}^\dagger &=  u_{\bm{k}}^* \alpha_{\bm{k}}^\dagger  +  v_{\bm{k}} \beta_{-\bm{k}}, \\
     b_{-\bm{k}} &=  v_{\bm{k}} \alpha_{\bm{k}}^\dagger  + u_{\bm{k}} \beta_{-\bm{k}}, \\
      b_{\bm{k}}^\dagger &= u_{\bm{k}}^* \beta_{\bm{k}}^\dagger  + v_{\bm{k}} \alpha_{-\bm{k}}, \\
      a_{-\bm{k}} &= v_{\bm{k}} \beta_{\bm{k}}^\dagger  + u_{\bm{k}} \alpha_{-\bm{k}}.
\end{align}
\label{eq:Bogoliubov}
\end{subequations}
We insert Eqs.~\eqref{eq:Bogoliubov} into $ H_3^D$, which now reads
\begin{align}
     H_3^D =&\, \sum_{\bm{k}, \bm{q}} \left[ D^+_{\bm{k}} \left( u^*_{\bm{k} + \bm{q}} \beta_{\bm{k}+ \bm{q}}^\dagger + v_{\bm{k} + \bm{q}} \alpha_{-\bm{k}- \bm{q}} \right) \left( v_{-\bm{k}} \beta_{-\bm{k}}^\dagger + u_{-\bm{k}} \alpha_{\bm{k}} \right) \left( v_{- \bm{q}} \alpha_{-\bm{q}}^\dagger + u_{- \bm{q}} \beta_{\bm{q}} \right) \right. \nonumber \\
      & \left. \qquad + D^+_{\bm{k}} \left( u^*_{\bm{k} } \beta_{\bm{k}}^\dagger + v_{\bm{k} } \alpha_{-\bm{k}} \right) \left( u^*_{\bm{q}} \alpha_{\bm{q}}^\dagger + v_{\bm{q}} \beta_{-\bm{q}} \right) \left( v_{- \bm{k} - \bm{q}} \beta_{-\bm{k}-\bm{q}}^\dagger + u_{- \bm{k} -\bm{q}} \alpha_{\bm{k}+\bm{q}} \right) \right. \nonumber \\
      & \left. \qquad+ D^-_{\bm{k}} \left( u^*_{\bm{k} } \alpha_{\bm{k}}^\dagger + v_{\bm{k}} \beta_{-\bm{k}} \right) \left( u^*_{\bm{q}} \beta_{\bm{q}}^\dagger + v_{\bm{q}} \alpha_{-\bm{q}} \right) \left( v_{- \bm{k}- \bm{q}} \alpha_{-\bm{k}-\bm{q}}^\dagger + u_{- \bm{k} - \bm{q}} \beta_{\bm{k}+ \bm{q}} \right) \right. \nonumber \\
      & \left. \qquad+ D^-_{\bm{k}} \left( u^*_{\bm{k} + \bm{q}} \alpha_{\bm{k}+ \bm{q}}^\dagger + v_{\bm{k} + \bm{q}} \beta_{-\bm{k}- \bm{q}} \right) \left( v_{-\bm{k}} \alpha_{-\bm{k}}^\dagger + u_{-\bm{k}} \beta_{\bm{k}} \right) \left( v_{- \bm{q}} \beta_{-\bm{q}}^\dagger + u_{- \bm{q}} \alpha_{\bm{q}} \right)  \right].
      \label{eq:H3innormalmodes}
\end{align}

\subsection{Lifting of spurious symmetries by three-magnon interactions}
Note that according to Eq.~\eqref{eq:H3innormalmodes} the three-magnon scattering is never spin conserving, because any combination of three normal modes, be it $\alpha^\dagger \alpha \beta$ or $\alpha \alpha \beta$ or any other, changes the spin. Thus, $H_3^D$ explicitly breaks the spurious spin conservation. It also breaks the spurious $C'_{2z}$ symmetry, because the three-magnon vertices contain $D^\pm_{\bm{k}}$, which itself contains
\begin{align}
    -\bm{D}_\delta \cdot 
    \left(
    \hat{\bm{y}} \pm \mathrm{i} \hat{\bm{x}}
    \right).
\end{align}
We can always choose the local tripod $(\hat{\bm{x}}, \hat{\bm{y}}, \hat{\bm{N}})$ so that $\hat{\bm{y}}$ is strictly in the crystallographic $ab$ (or $xy$) plane, that is $\hat{\bm{y}} = ( \hat{y}_x, \hat{y}_y, 0 )^\text{T}$. As a result, the third direction can be written as
\begin{align}
    \hat{\bm{x}} 
    =  
    \begin{pmatrix}
        \hat{y}_x \\ \hat{y}_y \\ 0
    \end{pmatrix}
    \times 
    \begin{pmatrix}
        \hat{N}_x \\ \hat{N}_y \\ \hat{N}_z
    \end{pmatrix}
    =
    \begin{pmatrix}
        \hat{y}_y \hat{N}_z \\  -\hat{y}_x \hat{N}_z \\ \hat{y}_x \hat{N}_y - \hat{y}_y \hat{N}_x
    \end{pmatrix},
\end{align}
resulting in
\begin{align}
    -\bm{D}_\delta \cdot 
    \left(
    \hat{\bm{y}} \pm \mathrm{i} \hat{\bm{x}}
    \right)
    =
    -\begin{pmatrix}
        D_{\delta,x} \\ D_{\delta,y} \\ 0
    \end{pmatrix}
    \cdot
    \begin{pmatrix}
        \hat{y}_x \pm \mathrm{i} \hat{y}_y \hat{N}_z \\  \hat{y}_y \mp \mathrm{i}\hat{y}_x \hat{N}_z \\ \pm \mathrm{i} (\hat{y}_x \hat{N}_y - \hat{y}_y \hat{N}_x)
    \end{pmatrix}
    =
        -D_{\delta,x} 
        \left(\hat{y}_x \pm \mathrm{i} \hat{y}_y \hat{N}_z \right) 
        + D_{\delta,y} \left(\hat{y}_y \pm \mathrm{i}\hat{y}_x \hat{N}_z \right).
\end{align}
Thus, the three-magnon vertices are explicitly dependent on $\hat{N}_z$ and, hence, capture the fact that $T C_{2z}$ is not a symmetry. Consequently, we expect that the inclusion of magnon-magnon interactions gives rise to a finite thermal Hall conductivity $\kappa_{xy}$ and spin Nernst conductivity $\alpha^\gamma_{xy}$.

\begin{figure}[]
       \includegraphics[width=15cm]{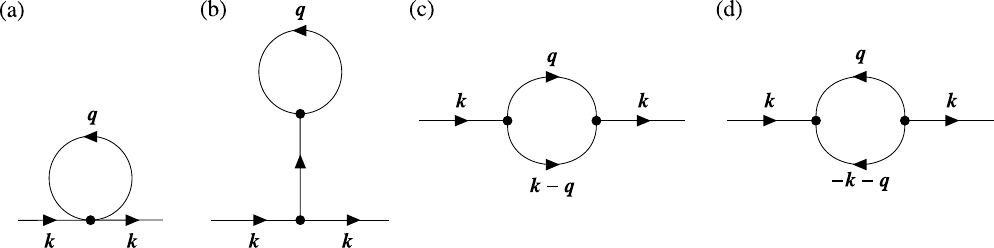}
       \caption{Feynman diagrams capturing leading $1/S$ corrections to the magnon spectrum at zero temperature. (a) Hartree diagram of four-magnon interactions. (b) Tadpole diagram of three-magnon interactions. (c) Forward bubble diagram of three-magnon interactions. (d) Backwards bubble diagram of three-magnon interactions.}
       \label{fig:feynman_diagram}
\end{figure}

In the rest of this section, we sketch what one would have to do to capture the symmetry lifting qualitatively. Within many-body perturbation theory, one can calculate the corrections to the magnon spectrum systematically in $1/S$. To leading order, four-magnon interactions enter in first-order perturbation theory and three-magnon interactions in second-order perturbation theory. The four-magnon interactions derive from the exchange interaction and correct the spectrum quantitatively via the Hartree diagram [see Fig.~\ref{fig:feynman_diagram}(a)], but they do not lift the spurious symmetries. The three-magnon interactions affect the spectrum by two different effects: (1) they can change the ground state by renormalizing the canting angle between the sublattices, thereby also changing the spectrum, and (2) they directly lead to scattering between the magnons. The first mechanism [see Fig.~\ref{fig:feynman_diagram}(b)], which may lead to a weak ferromagnetic moment, causes real corrections to the spectrum, but no lifetime effects.
The second mechanism has two contributions at zero temperature, which are shown diagrammatically in Figs.~\ref{fig:feynman_diagram}(c) and (d). Below, we discuss a strategy to see how these diagrams lead to a magnon-magnon hybridization, concentrating on the ``backwards bubble''. 

Starting from Eq.~\eqref{eq:H3innormalmodes}, we only retain source and sink terms (three magnons getting created or destroyed) and find
    \begin{align}
         H_3^{D, \text{source}} 
           =&\, \sum_{\bm{k}, \bm{q}, \bm{p}} \left( W^+_{\bm{k}, \bm{q}, \bm{p}}  \beta_{\bm{k}}^\dagger \alpha_{\bm{q}}^\dagger \beta_{\bm{p}}^\dagger +  W^-_{\bm{k}, \bm{q}, \bm{p}}  \alpha_{\bm{k}}^\dagger \beta_{\bm{q}}^\dagger \alpha_{\bm{p}}^\dagger + \mathrm{H.c.}\right).
    \end{align}
We used $D^\pm_{\bm{k}} = D^\pm_{-\bm{k}}$, as well as the definition
\begin{align}
     W^\pm_{\bm{k}, \bm{q}, \bm{p}} = D^\pm_{\bm{k}}  \left( v_{\bm{k}} v_{\bm{q}}  u^*_{\bm{p}} + u^*_{\bm{k}} u^*_{\bm{q}} v_{\bm{p}} \right).
\end{align}
To symmetrize the vertex we use a new notation of the normal modes, identifying $\gamma_1 = \alpha$ and $\gamma_2 = \beta$, so that
\begin{align}
    H_3^{D, \text{source}} 
    &=  \frac{1}{6} \sum_{\bm{k}, \bm{q}, \bm{p}}^{\bm{p} = - \bm{k} - \bm{q}} \sum_{\lambda, \mu, \nu} \left( \mathcal{W}_{\bm{k}, \bm{q}, \bm{p}}^{\lambda \mu \nu} \gamma_{\lambda, \bm{k}}^\dagger  \gamma_{\mu, \bm{q}}^\dagger  \gamma_{\nu, \bm{p}}^\dagger + \mathrm{H.c.}\right)
\end{align}
(where $\lambda$, $\mu$, and $\nu$ can assume values $1$ or $2$), with 
\begin{align}
    \mathcal{W}_{\bm{k}, \bm{q}, \bm{p}}^{\lambda \mu \nu} = W_{\bm{k}, \bm{q}, \bm{p}}^{\lambda \mu \nu} + 
    W_{\bm{q}, \bm{p}, \bm{k}}^{\mu \nu \lambda }  + W_{\bm{p}, \bm{k}, \bm{q}}^{\nu \lambda \mu } +  W_{\bm{k}, \bm{p}, \bm{q}}^{\lambda  \nu \mu} + W_{\bm{p}, \bm{q}, \bm{k}}^{ \nu \mu \lambda } + W_{\bm{q}, \bm{k}, \bm{p}}^{\mu \lambda  \nu}
\end{align}
and
\begin{align}
    W_{\bm{k}, \bm{q}, \bm{p}}^{\lambda \mu \nu} = 
    \begin{cases}
     W^{212}_{\bm{k}, \bm{q}, \bm{p}} = W^+_{\bm{k}, \bm{q}, \bm{p}} \quad \text{for } \lambda = 2, \mu = 1, \nu = 2 \\
     W^{121}_{\bm{k}, \bm{q}, \bm{p}}  = W^-_{\bm{k}, \bm{q}, \bm{p}} \quad \text{for } \lambda = 1, \mu = 2, \nu = 1 \\
     0 \qquad \quad \text{otherwise}
    \end{cases}.
    \label{eq:Ws}
\end{align}

The self-energies for the source and sink terms are computed as \cite{zhitomirskychernyshev2013} 
\begin{align}
    \Sigma_{\bm{k}}^{\alpha \beta} (\varepsilon) 
    = 
    - \frac{1}{2} \sum_{\bm{q}} \sum_{\alpha',\beta'} \frac{\mathcal{W}^{\alpha \alpha' \beta'}_{\bm{k}, \bm{q}, -\bm{k}-\bm{q}} \left( \mathcal{W}^{\beta \alpha' \beta'}_{\bm{k}, \bm{q}, -\bm{k}-\bm{q}}\right)^*}{\varepsilon - \mathrm{i}0^+ + E_{\bm{q},\alpha'} +  E_{ -\bm{k}-\bm{q}, \beta'}}.
\end{align}
We find the finite terms 
\begin{subequations}
    \begin{align}
        \Sigma_{\bm{k}}^{11}(\varepsilon) &= -\frac{1}{2} \sum_{\bm{q}} \left[ \frac{ \mathcal{W}_{\bm{k}, \bm{q},  -\bm{k}-\bm{q}}^{112} \left( \mathcal{W}_{\bm{k}, \bm{q},  -\bm{k}-\bm{q}}^{112}\right)^*}{\varepsilon + E_{ \bm{q}, 1} +  E_{-\bm{k}-\bm{q}, 2}}  
        + \frac{ \mathcal{W}_{\bm{k}, \bm{q},  -\bm{k}-\bm{q}}^{121} \left( \mathcal{W}_{\bm{k}, \bm{q},  -\bm{k}-\bm{q}}^{121}\right)^*}{\varepsilon + E_{\bm{q}, 2} +  E_{-\bm{k}-\bm{q}, 1}}  
        + \frac{ \mathcal{W}_{\bm{k}, \bm{q},  -\bm{k}-\bm{q}}^{122} \left( \mathcal{W}_{\bm{k}, \bm{q},  -\bm{k}-\bm{q}}^{122}\right)^*}{\varepsilon + E_{\bm{q}, 2} +  E_{-\bm{k}-\bm{q}, 2}} \right], \\
        \Sigma_{\bm{k}}^{12}(\varepsilon) &= -\frac{1}{2} \sum_{\bm{q}} \left[ \frac{ \mathcal{W}_{\bm{k}, \bm{q},  -\bm{k}-\bm{q}}^{112} \left( \mathcal{W}_{\bm{k}, \bm{q},  -\bm{k}-\bm{q}}^{212}\right)^*}{\varepsilon + E_{\bm{q}, 1} +  E_{-\bm{k}-\bm{q}, 2}}  
        + \frac{ \mathcal{W}_{\bm{k}, \bm{q},  -\bm{k}-\bm{q}}^{121} \left( \mathcal{W}_{\bm{k}, \bm{q},  -\bm{k}-\bm{q}}^{221}\right)^*}{\varepsilon + E_{\bm{q}, 2} +  E_{-\bm{k}-\bm{q}, 1}}  \right], \\
        \Sigma_{\bm{k}}^{21}(\varepsilon) &= -\frac{1}{2} \sum_{\bm{q}} \left[ \frac{ \mathcal{W}_{\bm{k}, \bm{q},  -\bm{k}-\bm{q}}^{212} \left( \mathcal{W}_{\bm{k}, \bm{q},  -\bm{k}-\bm{q}}^{112}\right)^*}{\varepsilon + E_{ \bm{q}, 1} +  E_{-\bm{k}-\bm{q}, 2}}  
        + \frac{ \mathcal{W}_{\bm{k}, \bm{q},  -\bm{k}-\bm{q}}^{221} \left( \mathcal{W}_{\bm{k}, \bm{q},  -\bm{k}-\bm{q}}^{121}\right)^*}{\varepsilon + E_{\bm{q}, 2} +  E_{-\bm{k}-\bm{q}, 1}}  \right], \\
        \Sigma_{\bm{k}}^{22}(\varepsilon) &= -\frac{1}{2} \sum_{\bm{q}} \left[ \frac{ \mathcal{W}_{\bm{k}, \bm{q},  -\bm{k}-\bm{q}}^{211} \left( \mathcal{W}_{\bm{k}, \bm{q},  -\bm{k}-\bm{q}}^{211}\right)^*}{\varepsilon + E_{ \bm{q}, 1} +  E_{-\bm{k}-\bm{q}, 1}}  
        + \frac{ \mathcal{W}_{\bm{k}, \bm{q},  -\bm{k}-\bm{q}}^{212} \left( \mathcal{W}_{\bm{k}, \bm{q},  -\bm{k}-\bm{q}}^{212}\right)^*}{\varepsilon + E_{\bm{q}, 1} +  E_{-\bm{k}-\bm{q}, 2}}  
        + \frac{ \mathcal{W}_{\bm{k}, \bm{q},  -\bm{k}-\bm{q}}^{221} \left( \mathcal{W}_{\bm{k}, \bm{q},  -\bm{k}-\bm{q}}^{221}\right)^*}{\varepsilon + E_{\bm{q}, 2} +  E_{ -\bm{k}-\bm{q}, 1}} \right],
    \end{align}
\end{subequations}
with [cf. Eq.~\eqref{eq:Ws}]
\begin{subequations}
\begin{align}
    \mathcal{W}_{\bm{k}, \bm{q}, \bm{p}}^{112} = W_{\bm{q}, \bm{p}, \bm{k}}^{121}  + W_{\bm{k}, \bm{p}, \bm{q}}^{121}, \\
    \mathcal{W}_{\bm{k}, \bm{q}, \bm{p}}^{121} = W_{\bm{k}, \bm{q}, \bm{p}}^{121}  + W_{\bm{p}, \bm{q}, \bm{k}}^{121} ,\\
     \mathcal{W}_{\bm{k}, \bm{q}, \bm{p}}^{211} = W_{\bm{p}, \bm{k}, \bm{q}}^{121}  + W_{\bm{q}, \bm{k}, \bm{p}}^{121}, \\
    \mathcal{W}_{\bm{k}, \bm{q}, \bm{p}}^{122} = W_{\bm{p}, \bm{k}, \bm{q}}^{212}  + W_{\bm{q}, \bm{k}, \bm{p}}^{212} ,\\
    \mathcal{W}_{\bm{k}, \bm{q}, \bm{p}}^{212} = W_{\bm{k}, \bm{q}, \bm{p}}^{212}  + W_{\bm{p}, \bm{q}, \bm{k}}^{212} ,\\
    \mathcal{W}_{\bm{k}, \bm{q}, \bm{p}}^{221} = W_{\bm{q}, \bm{p}, \bm{k}}^{212}  + W_{\bm{k}, \bm{p}, \bm{q}}^{212}.
\end{align}
\end{subequations}

The such obtained self-energy can be used to define a new effective harmonic Hamiltonian. Assuming small $\Delta$, we may use the on-shell approximation, with $\varepsilon = \varepsilon_{\bm{k}}$ as given in Eq.~\eqref{eq:energies}. In the band basis, the effective Hamiltonian reads
\begin{align}
    H_2^{\text{eff}} = \sum_{\bm{k}} 
    \begin{pNiceMatrix}
        \alpha_{\bm{k}}^\dagger & \beta_{\bm{k}}^\dagger
    \end{pNiceMatrix} H^{\text{eff}}_{\bm{k}} 
    \begin{pNiceMatrix}
        \alpha_{\bm{k}} \\ \beta_{\bm{k}}
    \end{pNiceMatrix},
    \label{eq:eff-Hamiltonian}
\end{align}
where
\begin{align}
    H^{\text{eff}}_{\bm{k}} 
    =
    \begin{pNiceMatrix}
        \varepsilon_{\bm{k}} + \Delta_{\bm{k}} & 0 \\
        0 & \varepsilon_{\bm{k}} - \Delta_{\bm{k}}
    \end{pNiceMatrix} + 
    \begin{pNiceMatrix}
        \Sigma_{\bm{k}}^{11} (\varepsilon_{\bm{k}}) &  \Sigma_{\bm{k}}^{12} (\varepsilon_{\bm{k}}) \\
         \Sigma_{\bm{k}}^{21} (\varepsilon_{\bm{k}}) &  \Sigma_{\bm{k}}^{22} (\varepsilon_{\bm{k}})
    \end{pNiceMatrix}.
\end{align}
Here, the first matrix contains the diagonalized harmonic spectrum and the second matrix contains (and approximation to) the self-energy. Importantly, the latter is not diagonal, because it is not spin conserving. As a result, a diagonalization of $ H^{\text{eff}}_{\bm{k}}$ hybridizes the $\alpha$ and $\beta$ magnons, lifting band degeneracies.




\begin{figure}[]
       \includegraphics[width=12cm]{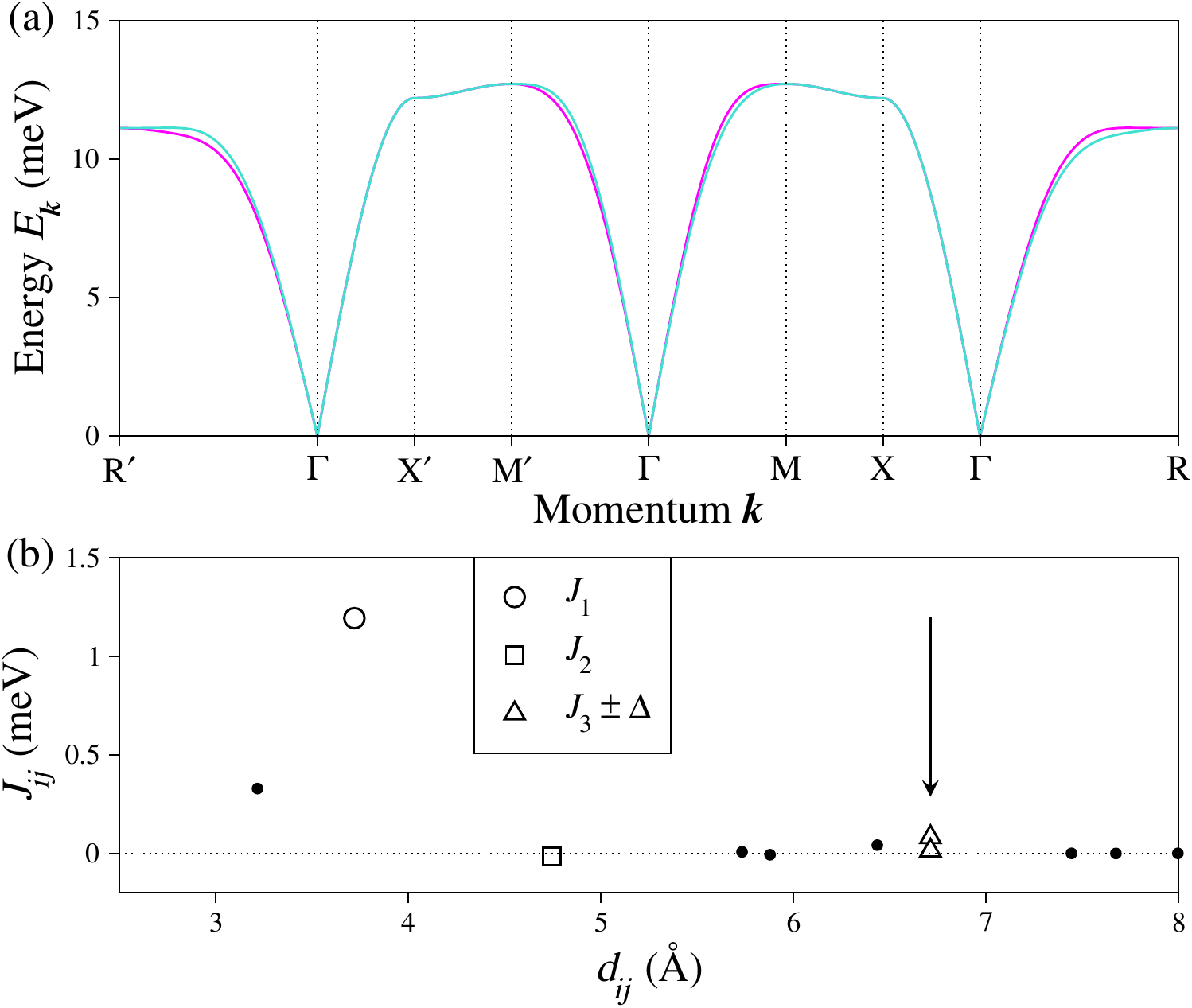}
       \caption{ \textit{Ab initio} results for the effective Heisenberg exchange parameters
and magnon spectrum of CoF$_2$ in the absence of spin-orbit coupling.
(a) Magnon dispersion relation of CoF$_2$.
(b) Corresponding Heisenberg exchange constants $J_{ij}$ as a function
of the distance $d_{ij}$ between magnetic sites.
The black arrow indicates the different exchange constants responsible
for the altermagnetic splitting.
The parameters read $S$ = 3/2, $J_1$ = 1.194 meV, $J_2$ = $-$0.016 meV,
$J_3$ = 0.048 meV, and $\Delta$ = 0.034 meV.}
       \label{fig:CoF2}
\end{figure}

\section{Proposed material candidates}
Our toy model is inspired by the rutile crystal structure, which is known to be realized, for example, in the insulating magnets MnF$_2$, NiF$_2$, and CoF$_2$. To the best of our knowledge, an altermagnetic magnon splitting has not been reported in experiments for any of these materials, prompting us to explore the splitting by \textit{ab initio} methods.
Following the methodology of Ref.~\onlinecite{SmejkalMagnons23}, we extract the Heisenberg
exchange constants for CoF$_2$ in the absence of spin-orbit coupling.
The generalized gradient approximation (GGA) \cite{Perdew1996} is used
with the projector-augmented wave (PAW) potentials \cite{Kresse1999}
implemented in {\sc vasp} software \cite{Kresse1996a,Kresse1996b}.
The lattice parameters are given by $a$ = 4.7468 $\mathring{\text{A}}$, $c$ = 3.2186 $\mathring{\text{A}}$,
and (0.3043, 0.3043, 0) for the F sites.
The tight-binding model is constructed by {\sc wannier90} code
\cite{Mostofi2014} including the Co-$d$ and F-$p$ orbitals as a basis set.
The magnetic exchange parameters for neighboring Co pairs are extracted
by {\sc tb2j} package \cite{He2021} based on the local magnetic force
theorem \cite{Liechtenstein1984}.
We calculate the magnon dispersion and identify the characteristic spin
or chirality splitting of the magnon bands [cf.~Figs~\ref{fig:CoF2}(a)].
Indeed, the magnon splitting is found to be tiny, albeit nonzero.
For reference, we plot the extracted Heisenberg exchange parameters
$J_{ij}$ in Fig.~\ref{fig:CoF2}(b) as a function of the distance $d_{ij}$
between magnetic atoms $i$ and $j$.
It is obvious that only the nearest and second-nearest-neighbor
interactions are sizable.
Responsible for the alternating splitting are the two different exchange
constants at~$d_{ij}$~=~$\sqrt{2}a$~=~6.713~$\mathring{\text{A}}$ [indicated by an arrow
in Fig.~\ref{fig:CoF2}(b)], which are rather small.

As we have established, the THE can be present even in the absence of the altermagnetic splitting. Thus, the absence of the splitting does not rule out the THE, which we suggest to explore in these materials. Our symmetry analysis has shown that the THE is absent if the Néel vector points along the $z$ axis (or $c$ axis), as is the case for CoF$_2$ and MnF$_2$. Therefore, we believe that NiF$_2$, for which the Néel vector is in the $xy$ plane (or $ab$ plane) \cite{Moriya1960NiF2}, is the most promising candidate to detect a THE in insulating altermagnets with a rutile crystal structure. We emphasize, however, that additional theoretical modelling is necessary to capture the THE quantitatively because NiF$_2$ is known to have local anisotropy axes \cite{Moriya1960NiF2}, which are neglected in our toy model.

\bibliography{bib}